\documentclass{article}
\usepackage{epsfig}
\usepackage{amsmath}
\usepackage{amsfonts}
\usepackage{amssymb}
\usepackage{euscript}
\usepackage{graphicx}
\usepackage{psfrag}
\usepackage{rotating}
\begin{document}
%\end{document}

\sloppy

%\small
%\pagestyle{plain}
\oddsidemargin=30mm
\evensidemargin=5mm

\hyphenpenalty=10000

$\qquad$

\vspace{20mm}
{\Large\underline{Annotation}}\\

Currently there is a common belief that the explanation of superconductivity phenomenon lies in understanding the mechanism of the formation of electron pairs.
Paired electrons, however, cannot form a superconducting condensate spontaneously. These paired electrons perform disorderly zero-point oscillations and there are no force of attraction in their ensemble. In order to create a unified ensemble of particles, the pairs must order their zero-point  fluctuations so that an attraction between the particles appears. As a result of this ordering of zero-point oscillations in the electron gas, superconductivity arises.  This model of condensation of zero-point oscillations creates the possibility of being able to obtain estimates for the critical parameters of elementary superconductors, which are in satisfactory agreement with the measured data.  On the another hand, the phenomenon of superfluidity in He-4 and He-3 can be similarly explained, due to the ordering of zero-point fluctuations. It is therefore established that both related phenomena are based on the same physical mechanism.\\

%PACS74.20.-z = theories and models of superconductivity

%PACS74.20.Mn  = nonconventional  mechanisms in superconductivity

\newpage

%\large

\thispagestyle{empty}

%\newpage

\begin{center}
{
\Large

\itshape{ Boris V.Vasiliev}}
\end{center}
\vspace{30mm}

\sloppy

\begin{center}
{{\Huge \bfseries SUPERCONDUCTIVITY\\
\vspace{20mm}
and\\
\vspace{20mm}
SUPERFLUIDITY
}}
\end{center}
%\vspace{-7cm}
\newpage

\tableofcontents

\newpage
\section{Superconductivity  and zero-point oscillations}

\subsection[Superconductivity and ordering of zero-point oscillations]{Superconductivity as a consequence of ordering of zero-point oscillations in electron gas}

\subsubsection[Superconductivity and superfluidity]{Superconductivity and superfluidity}
Superfluidity and superconductivity, which can be regarded as the superfluidity of the electron gas, are related phenomena.
The main feature of these phenomena can be seen in a fact that a special condensate  in superconductors as well as in superfluid helium   is formed from particles  interconnected by attraction.
This mutual attraction does not allow a scattering of individual particles  on defects and walls, if the energy of this scattering is less than the energy of attraction.
Due to the lack of scattering condensate acquires ability to move without friction.

Superconductivity was discovered over a century ago, and the superfluidity  about thirty years later.

However, despite the attention of many scientists to the study of these phenomena, they have been the great mysteries in condensed matter physics  for a long time.
This mystery attracted  the best minds of the twentieth century.

The mystery of the superconductivity phenomenon   has begun to drop in the middle of the last century when  the effect of magnetic flux quantization in superconducting cylinders was discovered and investigated.
This phenomenon was predicted even before the WWII by brothers F. London and H. London, but its quantitative study were  performed only two decades later.

By these measurements it became clear that at the formation of the superconducting state, two free electrons are combined into a single boson with zero spin and zero pulse.

Around the same time,  it was observed that the substitution of one isotope of the superconducting element to another leads to a changing of the critical temperature of superconductors: the phenomenon  called an isotope-effect \cite{Maxwell}, \cite{Serin}. This effect was interpreted as the direct proof of the key role of phonons in the formation of the superconducting state.

Following these understandings, L. Cooper proposed the phonon  mechanism of electron pairing on which base the microscopic theory of superconductivity (so called BCS-theory) was built by N. Bogolyubov and J. Bardin, L. Cooper and  J. Shriffer (probably it should be named better the Bogolyubov-BCS-theory).

However the B-BCS theory based on the phonon mechanism brokes a hypothetic link between superconductivity and superfluidity as in liquid helium there are no phonons for combining atoms.

Something similar happened with the description of superfluidity.

Soon after discovery of superfluidity, L.D. Landau in his first papers on the subject immediately demonstrated that this phenomenon should be considered as a result of condensate formation  consisting of macroscopic number of atoms in the same quantum state and obeying quantum laws.
It gave the possibility  to describe the main features of this phenomenon: the temperature dependence of the superfluid phase density, the existence of the second sound, etc.
But it does not gave an answer to the question which physical mechanism leads to the unification of the atoms in the superfluid condensate and what is the critical temperature of the condensate, i.e. why the ratio of the temperature of transition to the superfluid state to the boiling point of helium-4 is almost exactly equals to $1/2$, while for helium-3, it is about a thousand times smaller.

On the whole, the description of both super-phenomena, superconductivity and superfluidity, to the beginning of the twenty first century induced some feeling of dissatisfaction primarily due to the fact that  a common mechanism of their occurrence has not been understood.

More than fifty years of a study  of the B-BCS-theory has shown that this theory successfully describes the general features of the phenomenon, but it can not be
developed in the theory of superconductors.
It explains general laws such as the emergence of the energy gap, the behavior of  specific heat capacity, the flux quantization, etc.,
but it can not predict the main parameters of the individual superconductors: their critical temperatures and critical magnetic fields.
More precisely, in the B-BCS-theory, the expression for the critical temperature of superconductor obtains an exponential form which exponential factor is impossible to measure directly and  this formula is of no practical interest.

Recent studies of the isotopic substitution showed that zero-point oscillations of the ions in the metal lattice are not harmonical. Consequently the isotopic substitution affects the interatomic distances in a lattice, and as the result, they directly change  the Fermi energy of a metal  \cite{Inyu}.

 Therefore, the assumption developed in
the middle of the last century, that the electron-phonon interaction is the only possible mechanism of superconductivity was  proved to be wrong.
 The direct effect of isotopic substitution on the Fermi energy gives a possibility to  consider the superconductivity without the phonon mechanism.

Furthermore, a closer look  at the problem reveals that the B-BCS-theory describes the mechanism of electron pairing, but in this theory there is no mechanism for combining pairs in the single super-ensemble.
The necessary condition for the existence of superconductivity is  formation of a unique ensemble of particles.
By this mechanism, a very small amount of electrons are combined in super-ensemble, on the level 10 in minus fifth power from the full number of free electrons.
This fact also can not be understood in the framework of the B-BCS theory.

At very low temperatures, that allow superfluidity in helium and superconductivity in metals,  all movements of particles are freezed except for their zero-point oscillations.
Therefore, as an alternative, we should consider the interaction of super-particles through electro-magnetic fields of zero-point oscillations.
This approach was proved to be fruitful.
At the consideration of super-phenomena as  consequences of the zero-point oscillations ordering, one can construct theoretical mechanisms  enabling  to give  estimations for the critical parameters of these  phenomena which are in satisfactory agreement with measurements.

As result, one can see that as the critical temperatures of (type-I) superconductors are equal to about $10^{-6}$ from
the Fermi temperature for superconducting metal, which is consistent with data of measurements.
At this the destruction of superconductivity by application of critical magnetic field occurs when the field destroys the coherence of zero-point oscillations of electron pairs. This is  in good agreement with measurements also.

A such-like  mechanism works in superfluid liquid helium.
The problem of the interaction of zero-point oscillations of the electronic shells of neutral atoms in the s-state, was considered yet before the WWII by F.London.
He has shown that this interaction is responsible for the liquefaction of helium.
The closer analysis of  interactions of zero-point oscillations  for helium atomic shells shows that at first at the temperature of about 4K only one of the oscillations mode  becomes ordered. As a result, the forces of attraction appear between  atoms which are need for helium liquefaction.
To create a single quantum ensemble, it is necessary to reach the complete ordering of atomic oscillations.
At  the complete ordering of oscillations at about 2K, the additional energy of the mutual attraction appears and the system of helium-4 atoms transits in superfluid state.
To form the superfluid quantum ensemble in Helium-3, not only  the zero-point oscillations should be ordered, but   the magnetic moments of the nuclei should be ordered too.
For this reason, it is necessary to lower the temperature below 0.001K. This is also in agreement with experiment.

Thus it is possible to show that  both related super-phenomena, superconductivity and superfluidity, are based on the single physical mechanism: the ordering of  zero-point oscillations.

The roles of zero-point oscillations in formation of the superconducting state
have been previously considered in papers \cite{BV1}-\cite{BV3}.

\subsubsection{The electron pairing}
J.Bardeen was first who turned his attention toward a possible link between superconductivity and  zero-point oscillations  \cite{Bar}.

The special role of zero-point vibrations exists due to the fact that at low temperatures  all movements of electrons in metals have been  freezen except for these oscillations.

\begin{figure}
\hspace{-3cm}
\vspace{3cm}
\includegraphics[scale=0.7]{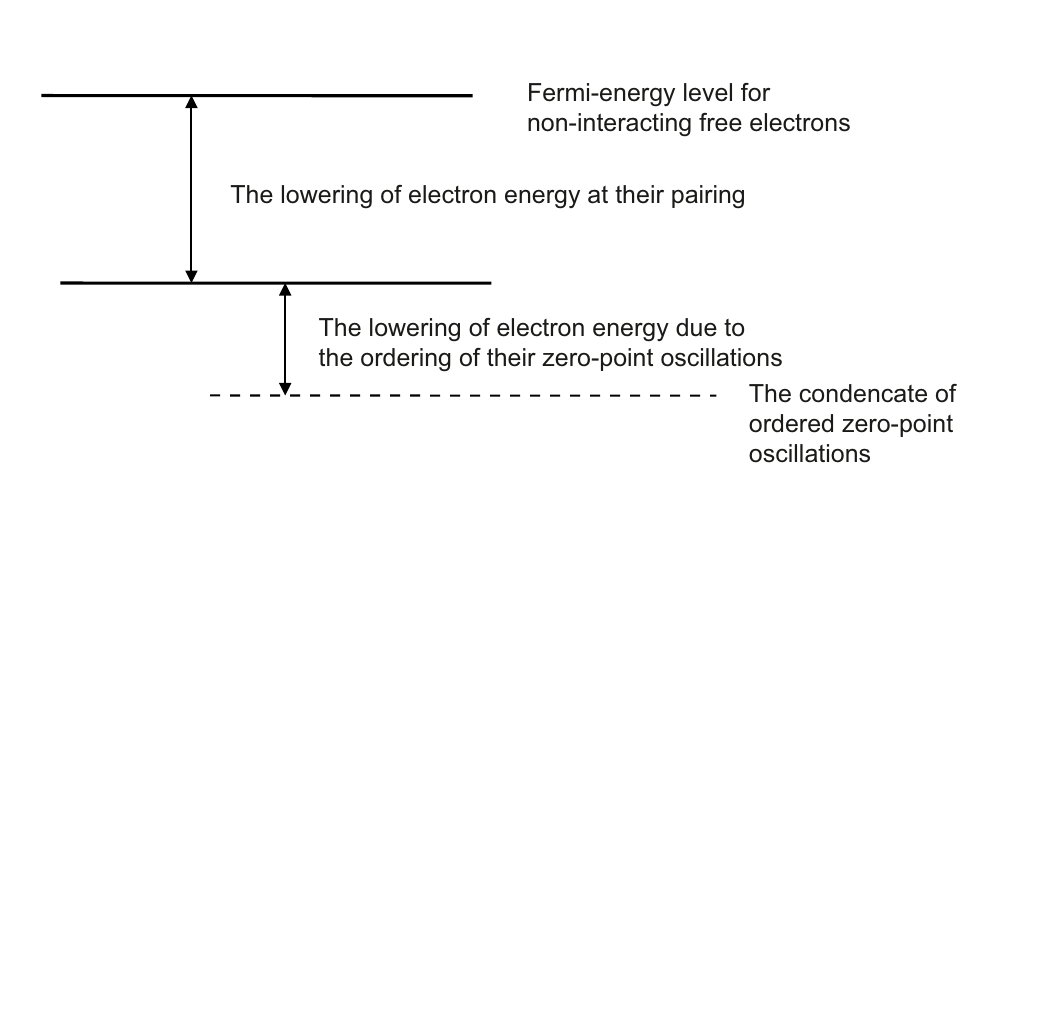}
\vspace{-8cm}\caption {The schematic representation of the energy levels of conducting electrons in a superconducting metal}\label{spe}
\end{figure}

Superconducting condensate formation requires two mechanisms:
 first, the electrons must be united in boson pairs, and then the zero-point fluctuations must be ordered (see Fig.(\ref{spe})).

The energetically favorable pairing of electrons in the electron gas should occur above the critical temperature.

Possibly, the pairing of electrons can occur due to the magnetic dipole-dipole interaction.

 For the magnetic dipole-dipole interaction, to merge two electrons  into the singlet pair at the temperature of about 10K, the distance between these particles must be small enough:
\begin{equation}
r<(\mu_B^2/kT_c)^{1/3}\approx a_B,\label{L1}
\end{equation}
where $a_B=\frac{\hbar^2}{m_e e^2}$ is the Bohr radius.

That is,  two collectivized electrons must be localized in one lattice site volume.
It is  agreed that the superconductivity can occur only in metals with two collectivized
electrons per atom, and cannot exist in the monovalent alkali and noble metals.

It is easy to see that the presence of  magnetic moments on  ion sites  should interfere with the magnetic combination of electrons. This is confirmed by the experimental fact: as there are no  strong magnetic substances among superconductors, so adding of iron, for example, to traditional superconducting alloys always leads to a lower  critical temperature.

On the other hand, this magnetic coupling should not be destroyed at the critical temperature. The energy of interaction between two electrons, located near one lattice site, can be much greater.
This is confirmed by experiments  showing that throughout the period of the magnetic flux quantization, there is no change at the transition through the critical temperature of superconductor \cite{Shab}, \cite{Sharv}.

The outcomes of these experiments are evidence that the existence of the mechanism of electron pairing is a necessary but not a sufficient condition for the existence of superconductivity.

The magnetic mechanism of electronic pairing proposed above  can be seen as an assumption which is consistent with the measurement data and therefore needs a more detailed theoretic consideration and further refinement.

 On the other hand, this issue is not very important in the grander scheme, because the nature of the mechanism that causes electron pairing is not of a significant importance.
 Instead, it is important that there is a mechanism which converts the electronic gas into an ensemble of charged bosons with zero spin in the considered temperature range (as well as in a some range of temperatures above $T_c$).

 If the temperature is not low enough, the electronic pairs still exist but their zero-point oscillations  are disordered. Upon reaching the $T_c$, the interaction between  zero-point oscillations should cause their ordering  and therefore a  superconducting state is created.

%\subsection{The condensate of ordered zero-point oscillations of electron gas}
\subsubsection{The interaction of zero-point oscillations}
%\bf{The interaction zero-point oscillations}
%\bf{The zero-point oscillations amplitude.}
The principal condition for the superconducting state formation is the ordering of zero-point oscillations. It is realized
because the paired electrons obeying Bose-Einstein statistics attract each other.

The origin of this attraction can be explained as follows.

\begin{figure}
\hspace{1.5cm}
\includegraphics[scale=0.5]{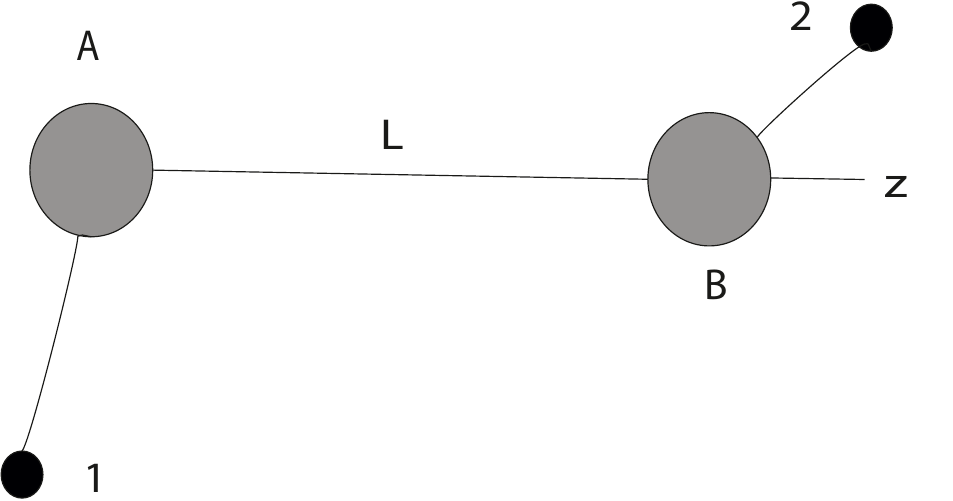}
\caption{Two ions placed on the distance $L$ and centers of their electronic clouds.}\label{v2}
\end{figure}

Let two  ions A and B be located on the z axis at the distance L from each other.
Two  collectivized electrons create clouds with centers at points 1 and 2 in the vicinity of each ions (Figure{\ref{v2}}).
Let $r_1$ be the radius-vector of the center of the first electronic cloud relative to the ion A and $ r_2 $ is the radius-vector of the second electron relative to the ion B.

Following the Born-Oppenheimer approximation, slowly oscillating ions are assumed fixed.
Let the temperature be low enough $ (T \rightarrow 0) $, so  only zero-point fluctuations of electrons would be taken into consideration.

In this case, the Hamiltonian of the system can be written as:
\begin{eqnarray}
H=H_0+H' \nonumber \\
H_0=-\frac{\hbar^2}{4m_e}\left(\nabla_1^2+\nabla_2^2\right)-\frac{4e^2}{r_1}-\frac{4e^2}{r_2}\\
H'=\frac{4e^2}{L}+\frac{4e^2}{r_{12}}-\frac{4e^2}{r_{1B}}-\frac{4e^2}{r_{2A}}\nonumber\label{LL1}
\end{eqnarray}
Eigenfunctions of the unperturbed Hamiltonian describes two ions surrounded by electronic clouds without interactions between them.
Due to the fact that the distance between the ions  is large compared with the size of the electron clouds $L\gg r$ , the additional term $H'$ characterizing the interaction can be regarded as a perturbation.

If we are interested in  the leading term of the interaction energy for L, the function  $ H '$ can be expanded in a series in powers of $1/L$ and we can write the first term:
{\begin{center}
\begin{equation}
\begin{array}{l}
{H'=\frac{4e^2}{L} \biggl\{1 +  }\\
%{H'=\frac{4e^2}{L} \left\{1 + \right. }\\
+\left[1+\frac{2(z_2-z_1)}{L}+\frac{(x_2-x_1)^2+(y_2-y_1)^2+(z_2-z_1)^2}{L^2}\right]^{-1/2}- \\
\left.-\left(1-\frac{2z_1}{L}+\frac{r_1^2}{L^2}\right)^{-1/2}-
\left(1+\frac{2z_2}{L}+\frac{r_2^2}{L^2}\right)^{-1/2} \right\}.
\end{array}
\end{equation}
\end{center}}
After combining the terms in this expression, we get:
\begin{equation}
H'\approx\frac{4e^2}{L^3}\left(x_1x_2+y_1y_2-2z_1z_2\right).\label{h1}
\end{equation}
This expression describes the interaction of two dipoles $ d_1 $ and $ d_2 $, which are formed by  fixed ions and electronic clouds of the corresponding instantaneous configuration.

Let us determine the displacements of  electrons which lead to an attraction in the system .

Let  zero-point fluctuations of the dipole moments formed by ions with their electronic clouds  occur with the  frequency $\Omega_0$, whereas each dipole moment can be decomposed into three orthogonal projection
$d_x=ex, d_y=ey $ and $d_z=ez$, and fluctuations of the second clouds are shifted in phase on $\varphi_x, \varphi_y $ and $\varphi_z$ relative to
fluctuations of the first.

As can be seen from Eq.(\ref{h1}), the interaction of z-components  is advantageous at in-phase  zero-point oscillations of  clouds,  i.e., when $\varphi_z = 2 \pi$.

Since the interaction of oscillating electric dipoles is due to the occurrence of oscillating electric field generated by them, the phase shift on $2\pi$ means that attracting  dipoles are placed  along the z-axis on the  wavelength $\Lambda_0$:
\begin{equation}
L_z=\Lambda_0=\frac{c}{2\pi \Omega_0}.\label{Lz1}
\end{equation}
As follows from (\ref{h1}), the attraction of dipoles at the interaction of the x and y-component will occur if these oscillations are antiphase, i.e. if the dipoles are separated along these axes on the distance equals to half of the wavelength:
\begin{equation}
L_{x,y}=\frac{\Lambda_0}{2}=\frac{c}{4\pi \Omega_0}.\label{Lxy}
\end{equation}
In this case
\begin{equation}
H'=-{4e^2}\left(\frac{x_1x_2}{L_x^3}+\frac{y_1y_2}{L_y^3}+2\frac{z_1z_2}{L_z^3}\right).\label{Lz3}
\end{equation}
Assuming that the electronic clouds have isotropic oscillations with amplitude $ a_0 $ for each axis
\begin{equation}
x_1=x_2=y_1=y_2=z_1=z_2=a_0
\end{equation}
we obtain
\begin{equation}
H'=576\pi^3\frac{e^2}{c^3}\Omega_0^3 a_0^2.\label{hh}
\end{equation}

\subsubsection{The zero-point oscillations amplitude}
The principal condition for the superconducting state formation, that is the ordering of zero-point oscillations, is realized
due to the fact that the paired electrons, which obey Bose-Einstein statistics, interact  with each other.

At they interact, their amplitudes, frequencies and phases of zero-point oscillations become ordered.

Let an electron gas has density $n_e$ and its Fermi-energy be $\mathcal{E}_F$. Each electron of this gas can be considered as fixed inside a cell with linear dimension $\lambda_F$:\footnote{Of course, the electrons are quantum particles and their fixation cannot be considered too literally. Due to the Coulomb forces of ions, it is more favorable for  collectivized electrons  to be placed near the ions  for the shielding  of ions fields. At the same time, collectivized electrons are spread over whole metal.
It is wrong to think that a particular electron is fixed inside a cell near to a particular ion.
But the spread of the electrons does not play a fundamental importance for our further consideration, since  there are two electrons near the node of the lattice in the divalent metal at any given time.
They  can be considered as located inside the cell as averaged.}
\begin{equation}
\lambda_F^3=\frac{1}{n_e}%\label{}
\end{equation}
which corresponds to the de Broglie wavelength:
\begin{equation}
\lambda_F=\frac{2\pi \hbar}{p_F}.\label{lF1}
\end{equation}
Having taken into account (\ref{lF1}), the Fermi energy of the electron gas can be written as
\begin{equation}
\mathcal{E}_F=\frac{p_F^2}{2m_e}=2\pi^2\frac{e^2a_B}{\lambda_F^2}.\label{EF2}
\end{equation}

However, a free electron interacts with the ion at its zero-point oscillations.
If we consider the  ions system as a positive background uniformly spread over the cells, the electron inside one cell  has the potential energy:
\begin{equation}
\mathcal{E}_p\simeq -\frac{e^2}{\lambda_F}.%\label{}
\end{equation}
As zero-point oscillations of the electron pair are   quantized by definition, their frequency and amplitude are related
\begin{equation}
{m_e a_0^2\Omega_0}\simeq \frac{\hbar}{2}.
\end{equation}
Therefore, the kinetic energy of electron undergoing zero-point oscillations in a limited region of space, can be written as:
\begin{equation}
\mathcal{E}_k\simeq \frac{\hbar^2}{2m_e a_0^2}.%\label{}
\end{equation}

In accordance with the virial theorem \cite{vir}, if a particle executes a finite motion,
its potential energy $\mathcal{E}_p$ should be associated with its kinetic energy $\mathcal{E}_k$ through the simple relation $|\mathcal{E}_p|=2\mathcal{E}_k$.

In this regard, we find that the amplitude of the zero-point oscillations of an electron in a cell is:
\begin{equation}
a_0\simeq \sqrt{2\lambda_F a_B}. \label{a0}
\end{equation}

\subsubsection{The condensation temperature}
Hence the interaction energy, which unites  particles into the condensate of ordered zero-point oscillations
\begin{equation}
\Delta_0\equiv H'=18\pi^3\alpha^3\frac{e^2a_B}{\lambda_F^2},\label{Dh1}
\end{equation}
where $\alpha=\frac{1}{137}$ is the fine structure constant.

Comparing this  association energy   with the Fermi energy (\ref{EF2}), we obtain
\begin{equation}
\frac{\Delta_0}{\mathcal{E}_F}= 9\pi \alpha^3\simeq 1.1\cdot 10^{-5}.\label{Dh2}
\end{equation}

Assuming that the critical temperature below which the possible existence of such condensate is approximately equal
\begin{equation}
T_c\simeq \frac{1}{2}\frac{\Delta_0}{k}% \label{a0}
\end{equation}
(the coefficient approximately equal to 1/2 corresponds to the experimental data, discussed below in the subsection (\ref{Delta-TcE})).

After substituting obtained parameters, we have
\begin{eqnarray}
T_c\simeq 5.5\cdot 10^{-6}{T}_F\label{TT0}
\end{eqnarray}

The experimentally measured ratios  $\frac{T_c}{T_F}$  for I-type superconductors are given in Table (\ref{de1}) and in Fig.(\ref{TT0g}).

The straight line on this figure is obtained from Eq.(\ref{TT0}), which
as seen defines an upper limit of critical temperatures of I-type superconductors.

\vspace{0.5cm}

\newpage

\subsection[The condensate and type-I superconductors]{The condensate of zero-point oscillations  and type-I superconductors}
\subsubsection{The critical temperature of type-I superconductors}
In order to compare the critical temperature of the condensate of zero-point oscillations with   measured critical temperatures of superconductors, at first we should make an estimation on the Fermi energies of superconductors.
For this we use the experimental data for the Sommerfeld`s constant  through which the Fermi energy can be expressed:
\begin{equation}
\gamma=\frac{\pi^2  k^2 n_e}{4\mathcal{E}_F}=\frac{1}{2}\cdot\left(\frac{\pi}{3}\right)^{2/3}\left(\frac{k}{\hbar}\right)^2 m_e n_e^{1/3}\label{gz}
\end{equation}
So on the basis of Eqs.(\ref{EF2}) and (\ref{gz}), we get:
\begin{equation}
kT_F(\gamma)=\frac{p_F^2(\gamma)}{2m_e}\simeq \left(\frac{12}{k^2}\right)^2\left(\frac{\hbar^2}{2m_e}\right)^3\gamma^2.\label{kTF}
\end{equation}

On base of these calculations we obtain possibility to relate  directly the critical temperature of a superconductor with the experimentally measurable parameter: with its electronic specific heat.

Taking into account Eq.(\ref{TT0}), we have:
\begin{equation}
\Delta_0\simeq \Theta\gamma^2\label{tcc},
\end{equation}
where the constant
\begin{equation}
\Theta\simeq 31\frac{\pi^2}{k} \left[\frac{\alpha\hbar^2}{k m_e}\right]^3\simeq 6.65\cdot 10^{-22}\frac{K^4 cm^6}{erg}\label{tc2}.
\end{equation}

\begin{figure}
\hspace{0.5cm}
\includegraphics[scale=0.5]{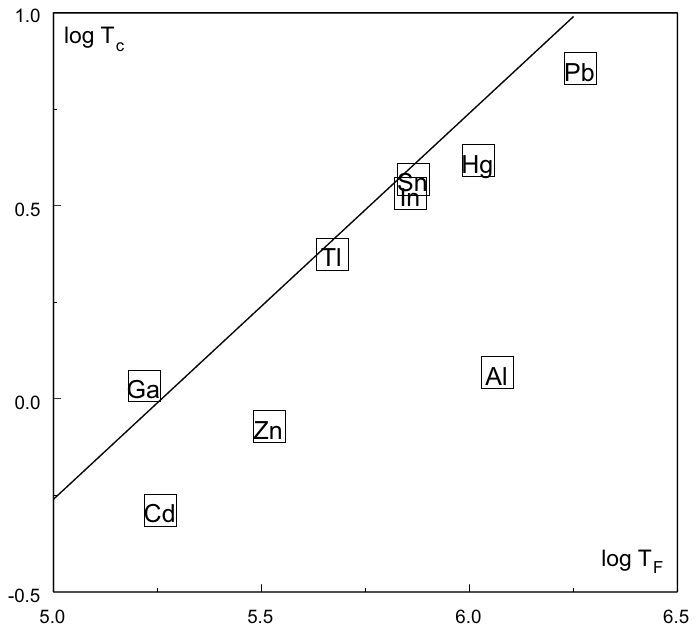}
\vspace{-7cm}\caption {The comparison of critical temperatures $T_c$ of type-I superconductors with their Fermi temperatures $T_F$. The straight line is obtained from Eq.(\ref{TT0}).}\label{TT0g}
\end{figure}

The comparison of the calculated parameters and  measured data ({\cite{Ketterson}},{\cite{Pool}})  is given in Table (\ref{de1})-(\ref{2tt}) and in Fig.({\ref{TT0g}}) and (\ref{tc2g}).\\
\bigskip

%\newpage
\begin{table}
\centering
\begin{tabular}{||c|c|c|c||}\hline\hline
&&&\\%\hline
  superconductor &$T_c$,K&$T_F$,K&$\frac{T_c}{T_F}$\\%\hline
  &&Eq(\ref{kTF})&\\
    &&&\\\hline
  Cd &0.51&$1.81\cdot 10^5$&$2.86\cdot 10^{-6}$\\
  Zn &0.85&$3.30\cdot 10^5$&$2.58\cdot 10^{-6}$\\
  Ga &1.09&$1.65\cdot 10^5$&$6.65\cdot 10^{-6}$\\
  Tl &2.39&$4.67\cdot 10^5$&$5.09\cdot 10^{-6}$\\
  In &3.41&$7.22\cdot 10^5$&$4.72\cdot 10^{-6}$\\
  Sn &3.72&$7.33\cdot 10^5$&$5.08\cdot 10^{-6}$\\
  Hg &4.15&$1.05\cdot 10^6$&$3.96\cdot 10^{-6}$\\
  Pb &7.19&$1.85\cdot 10^6$&$3.90\cdot 10^{-6}$\\ \hline\hline
  \end{tabular}
  \caption{The comparison of the calculated values of  superconductors critical temperatures with measured Fermi temperatures}
\label{de1}
\end{table}

\bigskip

\begin{table}
\centering
\begin{tabular}{||c|c|c|c|c||}\hline\hline
  super-&$T_c$(measur),&$\gamma,\frac{erg}{cm^3K^2}$&$T_c$(calc),K&$\frac{T_c (calc)}{T_c(meas)}$\\
conductors  &K&&Eq.(\ref{tcc})&\\\hline
  Cd &$0.517$&532&$0.77$&1.49\\
  Zn &$0.85$&718&$1.41$&1.65\\
  Ga &$1.09$&508&$0.70$&0.65\\
  Tl &$2.39$&855&$1.99$&0.84\\
  In &$3.41$&1062&$3.08$&0.90\\
  Sn &$3.72$&1070&$3.12$&0.84\\
  Hg &$4.15$&1280&$4.48$&1.07\\
  Pb &$7.19$&1699&$7.88$&1.09\\ \hline\hline
\end{tabular}
\caption{The comparison of the calculated values of superconductors critical temperatures with measurement data}
\label{2tt}
\end{table}

\bigskip

\bigskip

\bigskip

\subsubsection[The relation of critical parameters]{The relation of critical parameters of type-I superconductors}
\label{crit-param1}
The phenomenon of condensation of zero-point oscillations in the electron gas has its characteristic features.

 There are several ways of destroying  the zero-point oscillations condensate in electron gas:

 Firstly, it can be evaporated by heating. In this case, evaporation of the condensate should possess the properties of an order-disorder transition.

 Secondly, due to the fact that the oscillating electrons carry electric charge, the condensate can be destroyed by the application of a sufficiently strong magnetic field.

 For this reason, the critical temperature and critical magnetic field of the condensate will be interconnected.

  This interconnection should manifest itself through the relationship of the critical temperature and critical field of the superconductors, if superconductivity occurs as result of an ordering of zero-point fluctuations.

Let us assume that at a given temperature ${T <T_{c}}$ the system of vibrational levels of conducting electrons consists of only two levels:

 firstly, basic level which is characterized by an anti-phase oscillations of the electron pairs at the distance $\Lambda_0/2$, and

 secondly, an excited level characterized by in-phase oscillation of the pairs.

Let the population of the basic level be $N_0$ particles and the excited level has  $N_1$ particles.

Two electron pairs at an in-phase oscillations have a high energy of interaction and therefore cannot form the condensate.
The condensate can be formed only by the particles that make up the difference between the populations of levels $N_0-N_1$.
In a dimensionless form, this difference defines the order parameter:
\begin{equation}
\Psi=\frac{N_0}{N_0+N_1}-\frac{N_1}{N_0+N_1}.
\end{equation}
In the theory of superconductivity, by definition, the order parameter is determined by the value of the energy gap
\begin{equation}
\Psi=\Delta_T/\Delta_0.
\end{equation}
When taking a counting of energy from the level $\varepsilon_0$, we obtain
\begin{equation}
\frac{\Delta_{T}}{\Delta_0}=\frac{N_0-N_1}{N_0+N_1}\simeq\frac{e^{2\Delta_{T}/kT} -1}{e^{2\Delta_{T}/kT} +1}=th(2\Delta_{T}/kT).\label{det}
\end{equation}
Passing to dimensionless variables $\delta\equiv \frac{\Delta_{T}}{\Delta_0}$ , $t\equiv \frac{kT}{kT_c}$
and $\beta\equiv\frac{2\Delta_0}{kT_c}$ we have
\begin{equation}
\delta=\frac{e^{\beta\delta/t} -1}{e^{\beta\delta/t} +1}=th(\beta\delta/t).\label{del}
\end{equation}
This equation describes the temperature dependence of the energy gap in the spectrum of zero-point oscillations.
It is similar to other equations describing other physical phenomena, that are also characterized by the existence of the temperature dependence of order parameters \cite{LL},\cite{Kit}. For example, this dependence is similar to temperature dependencies of  the concentration of the superfluid component in liquid helium or the spontaneous magnetization of ferromagnetic materials. This equation is the same for all order-disorder transitions (the phase transitions of 2nd-type in the Landau classification).

\begin{figure}
\centering
\includegraphics[scale=0.75]{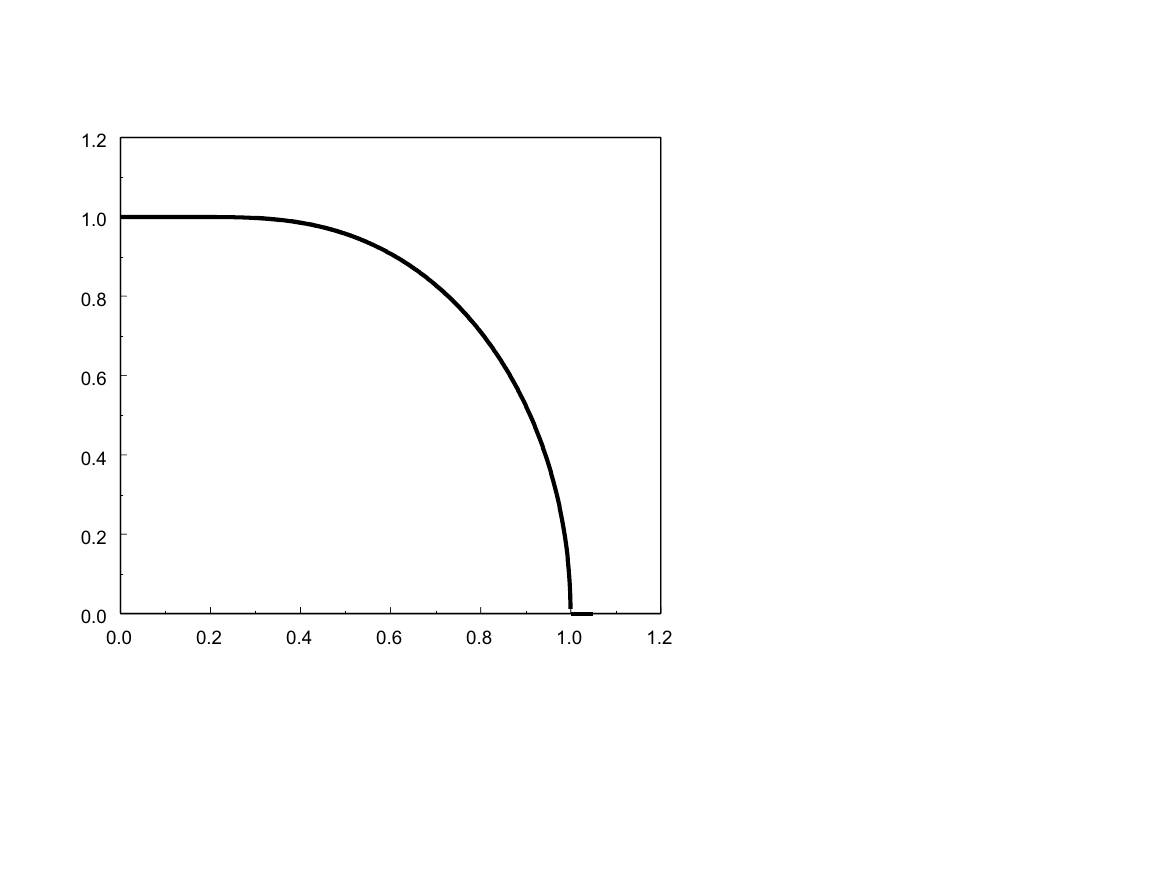}
\vspace{-3cm}\caption{The temperature dependence of the value of the gap in the energetic spectrum of zero-point oscillations calculated on Eq.(\ref{del}).}
\label{D-T}
\end{figure}

The solution of this equation, obtained by the iteration method, is shown in Fig.(\ref{D-T}).

This decision is in a agreement with  the known  transcendental equation of the BCS, which was obtained by the integration of the phonon spectrum, and is in a satisfactory agreement with the measurement data.

After numerical integrating we can obtain the averaging value of the gap:
\begin{equation}
\langle\Delta\rangle=\Delta_0\int_0^1 \delta dt=0.852~\Delta_0~.\label{0.8}
\end{equation}

To convert the condensate into the normal state, we  must  raise half of its particles  into the excited state  (according to Eq.(\ref{det}), the gap collapses under this condition). To do this, taking into account Eq.(\ref{0.8}), the unit volume of condensate   should have the energy:
\begin{equation}
\mathcal{E}_T\simeq  \frac{1}{2} n_0 \langle\Delta_0  \rangle  \approx  \frac{0.85}{2}\left(\frac{m_e}{2\pi^2\alpha\hbar^2}\right)^{3/2}\Delta_0^{5/2},\label{ET}
\end{equation}
On the other hand, we can obtain the normal state of an electrically charged condensate when applying a magnetic field of critical value $H_c$ with the density of energy:
\begin{equation}
\mathcal{E}_H= \frac{H_c^2}{8\pi}.\label{EH}
\end{equation}
As a result, we acquire the condition:
\begin{equation}
\frac{1}{2}n_0 \langle\Delta_0  \rangle=\frac{H_c^2}{8\pi}.\label{TH1}
\end{equation}
This creates a  relation of  the critical temperature to the critical magnetic field of the zero-point oscillations condensate of the charged bosons.

\begin{figure}
%\hspace{1.5cm}
\includegraphics[scale=0.5]{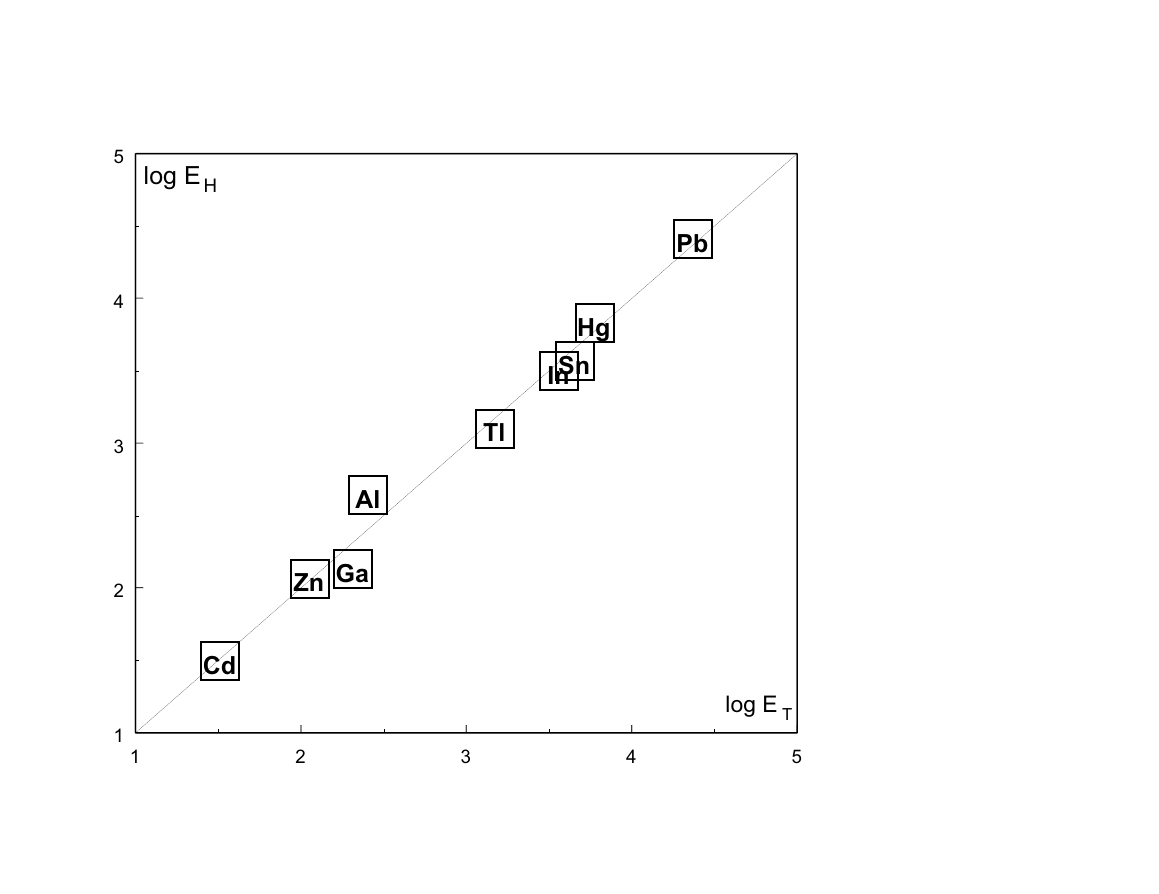}
\caption{The comparison of the critical energy densities $\mathcal{E}_T$ (Eq.(\ref{ET})) and $\mathcal{E}_H$ (Eq.(\ref{EH})) for the type-I superconductors.}\label{eh-et2}
\end{figure}

The comparison of the critical energy densities $\mathcal{E}_T$ and $\mathcal{E}_H$ for type-I superconductors are shown in Fig.(\ref{eh-et2}).
As shown,  the obtained agreement between  the energies $\mathcal{E_T}$ (Eq.(\ref{ET})) and $\mathcal{E_H}$ (Eq.(\ref{EH}))  is  quite satisfactory for type-I superconductors \cite{Ketterson},\cite{Pool}.
 A similar comparison for type-II superconductors shows results that differ by a factor two approximately.
 The reason for this will be considered below.
 The correction of this calculation, has not apparently made sense here.
 The purpose of these calculations was to show that the description of superconductivity as the effect of the condensation of ordered zero-point oscillations is in accordance with the available experimental data. This goal is considered reached in  the simple case of type-I superconductors.

\subsubsection[The critical magnetic field]{The critical magnetic field of superconductors}
The direct influence of the external magnetic field of the critical value applied to the electron system is too weak to disrupt the dipole-dipole interaction of two paired electrons:
\begin{equation}
\mu_B H_c \ll kT_c.\label{L2}
\end{equation}
 In order to violate the superconductivity so as to destroy the ordering of the electron zero-point oscillations. For this the presence of relatively weak magnetic field is required.

At combing of Eqs.(\ref{TH1}),(\ref{ET}) and (\ref{a0}), we can express the gap through the critical magnetic field and the magnitude of the oscillating dipole moment:
\begin{equation}
\Delta_0\approx \frac{1}{2}~e~a_0~H_c.\label{dah}
\end{equation}
The properties of the zero-point oscillations of the electrons should not be dependent on the characteristics of the mechanism of association and also on the condition of the existence of electron pairs. Therefore, we should expect that this equation would also be valid  for type-I superconductors, as well as for II-type superconductors (for II-type superconductor  $H_c=H_{c1}$ is the first critical field)

\begin{figure}
\vspace{-13.5cm}
\centering
\includegraphics[scale=0.8]{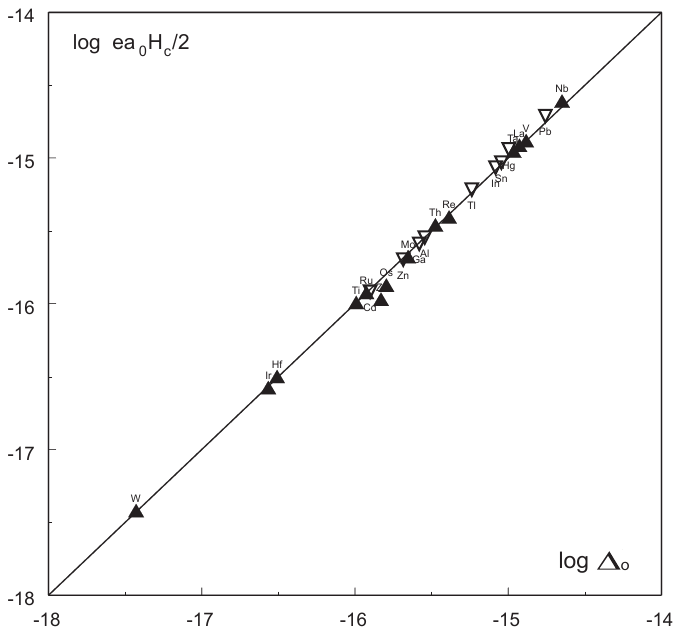}
\vspace{-2cm}\caption {The comparison of the calculated energy of superconducting pairs in the critical magnetic field with the value of the superconducting gap.
Here, the following key applies:
%\vspace{.1 cm}
filled triangles - type-II superconductors,
%\vspace{.1 cm}
empty triangles - type-I superconductors.
%\vspace{.1 cm}
On vertical axis - logarithm of the product of the calculated value of the oscillating  dipole moment of an electron pair on the critical magnetic field is plotted.%\vspace{.1 cm}
On horizontal axis - the value of the gap is shown.}\label{rd2}
\end{figure}

An agreement with this condition is illustrated on the Fig.(\ref{rd2}).

%\newpage
\bigskip

%\newpage

\subsubsection{The density of superconducting carriers}
Let us consider the process of heating the electron gas in metal.
 When heating, the electrons from levels slightly below the Fermi-energy are raised to higher levels. As a result, the levels closest to the Fermi level, from which at low temperature electrons were forming bosons, become vacant.

At critical temperature $T_c$, all electrons from the levels of energy bands from $\mathcal{E}_F-\Delta$ to  $\mathcal{E}_F$  move to higher levels (and the gap collapses). At this temperature superconductivity is therefore destroyed completely.

This band of energy can be filled by $N_\Delta$ particles:
\begin{equation}
N_\Delta=2\int_{\mathcal{E}_F-\Delta}^{\mathcal{E}_F}
F(\mathcal{E})D(\mathcal{E})d\mathcal{E}\label{ne}.
\end{equation}
Where  $F(\mathcal{E})=\frac{1}{e^{\frac{\mathcal{E}-\mu}{\tau}}+1}$ is the Fermi-Dirac function and $D(\mathcal{E})$ is  number of states per an unit energy interval, a deuce front of the integral arises from the fact that there are two electron at each energy level.

To find the density of states $D(\mathcal{E})$,  one needs to find the difference in energy of the system at $T=0$ and finite temperature:
\begin{equation}
\Delta \mathcal{E} =\int_0^\infty F(\mathcal{E})\mathcal{E}D(\mathcal{E})d\mathcal{E}-\int_0^{\mathcal{E}_F} \mathcal{E}D(\mathcal{E})d\mathcal{E}\label{dE}.
\end{equation}
 For the calculation of the density of states $D(\mathcal{E})$, we must note that two electrons can be placed on each level. Thus, from the expression of the Fermi-energy Eq.(\ref{EF2})
we obtain
\begin{equation}
D({E}_F)=\frac{1}{2}\cdot\frac{dn_e}{d\mathcal{E}_F}=\frac{3n_e}{4\mathcal{E}_F}=\frac{3\gamma}{2k^2\pi^2},\label{d}
\end{equation}
where
\begin{equation}
\gamma=\frac{\pi^2  k^2 n_e}{4\mathcal{E}_F}=\frac{1}{2}\cdot\left(\frac{\pi}{3}\right)^{3/2}\left(\frac{k}{\hbar}\right)^2 m_e n_e^{1/3}\label{gzz}
\end{equation}
is the Sommerfeld constant
\footnote{It should be noted that because on each level two electrons can be placed, the expression for the Sommerfeld constant Eq.(\ref{gzz}) contains the additional factor $1/2$ in comparison with the usual formula in  literature \cite{Kit}}.

Using similar arguments, we can calculate the number of electrons, which populate the levels in the range from $\mathcal{E}_F-\Delta$ to $\mathcal{E}_F$. For an unit volume of material, Eq.(\ref{ne}) can be rewritten as:
\begin{equation}
n_\Delta=2kT\cdot D(\mathcal{E}_F)\int_{-\frac{\Delta_0}{kT_c}}^0  \frac{dx}{(e^x +1)}. %\label{e4}
\end{equation}

By supposing that for superconductors $\frac{\Delta_0}{kT_c}=1.76$, as a result of numerical integration we obtain
\begin{equation}
\int_{-\frac{\Delta_0}{kT_c}}^0  \frac{dx}{(e^x +1)}=\left[x-ln(e^x+1)\right]_{-1.76}^0\approx 1.22 . %\label{e4}
\end{equation}
Thus, the density of electrons, which throw up above the Fermi level in a metal at temperature $T = T_c$ is
\begin{equation}
n_e(T_c)\approx 2.44 \left(\frac{3\gamma}{k^2\pi^2}\right)kT_c.\label{net}
\end{equation}
Where the Sommerfeld constant $\gamma$ is  related to the volume unit  of the metal.

%\newpage
From Eq.(\ref{Lxy}) it follows
\begin{equation}
L_0\simeq\frac{\lambda_F}{\pi\alpha}\label{L0}
\end{equation}
 and this  forms the ratio of the condensate particle density  to the Fermi gas density:
\begin{equation}
\frac{n_0}{n_e}=\frac{\lambda_F^3}{L_0^3}\simeq\left(\pi\alpha\right)^3\simeq 10^{-5}.\label{n0-ne}
\end{equation}
When using these equations, we can find a linear dimension of localization for an electron pair:
\begin{equation}
L_0 =\frac{\Lambda_0}{2}\simeq  \frac{1}{\pi\alpha(n_e)^{1/3}}.\label{L}
\end{equation}
or, taking into account Eq.(\ref{a0}), we can obtain the relation between the density of particles in the condensate and the value of the energy gap:
\begin{equation}
\Delta_0\simeq 2\pi^2\alpha\frac{\hbar^2}{m_e}n_0^{2/3}\label{D-2}
\end{equation}
or
\begin{equation}
n_0=\frac{1}{L_0^3}=\left(\frac{m_e}{2\pi^2\alpha \hbar^2}\Delta_0\right)^{3/2}.\label{n-0}
\end{equation}
It should be noted that the obtained ratios for the zero-point oscillations condensate (of bose-particles)
differ from the corresponding expressions for the bose-condensate of particles, which can be obtained in many courses (see eg \cite{LL}). The expressions for the ordered condensate of zero-point  oscillations have an additional coefficient $\alpha$ on the right side of Eq.(\ref{D-2}).

\vspace{0.5cm}
The de Broglie wavelengths of  Fermi electrons expressed through the Sommerfelds constant
\begin{equation}
\lambda_F=\frac{2\pi \hbar}{p_F(\gamma)}\simeq\frac{\pi}{3}\cdot\frac{k^2 m_e}{\hbar^2 \gamma}\label{lF}
\end{equation}
are shown in Tab.\ref{n0e}.

In accordance with Eq.(\ref{L0}), which was obtained at the zero-point oscillations consideration,  the ratio $\frac{\lambda_F}{\Lambda_0}\simeq 2.3\cdot 10^{-2}$.

In connection with this ratio, the calculated ratio of the zero-point oscillations condensate density to the density of fermions in accordance with Eq.(\ref{n0-ne}) should be near to $10^{-5}$.

 It can be therefore be seen, that calculated estimations of the condensate parameters  are in satisfactory agreement with experimental data of superconductors.
{
\begin{table}
\centering
\begin{tabular}{||c|c|c|c|c||}\hline\hline
&&&&\\%\hline
  super-&$\lambda_F$,cm&$\Lambda_0$,cm&$\frac{\lambda_F}{\Lambda_0}$&
  $\frac{n_0}{n_e}=\left(\frac{\lambda_F}{\Lambda_0}\right)^3$\\%\hline
   conductor &Eq(\ref{lF})&Eq(\ref{Lxy})&&\\
   &&&&\\\hline
  Cd &$3.1\cdot 10^{-8}$&$1.18\cdot 10^{-6}$&$2.6\cdot 10^{-2}$&$1.8\cdot 10^{-5}$\\
  Zn &$2.3\cdot 10^{-8}$&$0.92\cdot 10^{-6}$&$2.5\cdot 10^{-2}$&$1.5\cdot 10^{-5}$\\
  Ga &$3.2\cdot 10^{-8}$&$0.81\cdot 10^{-6}$&$3.9\cdot 10^{-2}$&$6.3\cdot 10^{-5}$\\
  Tl &$1.9\cdot 10^{-8}$&$0.55\cdot 10^{-6}$&$3.4\cdot 10^{-2}$&$4.3\cdot 10^{-5}$\\
  In &$1.5\cdot 10^{-8}$&$0.46\cdot 10^{-6}$&$3.2\cdot 10^{-2}$&$3.8\cdot 10^{-5}$\\
  Sn &$1.5\cdot 10^{-8}$&$0.44\cdot 10^{-6}$&$3.4\cdot 10^{-2}$&$4.3\cdot 10^{-5}$\\
  Hg &$1.3\cdot 10^{-8}$&$0.42\cdot 10^{-6}$&$3.1\cdot 10^{-2}$&$2.9\cdot 10^{-5}$\\
  Pb &$1.0\cdot 10^{-8}$&$0.32\cdot 10^{-6}$&$3.1\cdot 10^{-2}$&$2.9\cdot 10^{-5}$\\ \hline\hline
  \end{tabular}
 \caption{The ratios $\frac{\lambda_F}{\Lambda_0}$ and $\frac{n_0}{n_e}$ for type-I superconductors}
\label{n0e}
\end{table}

%Table \ref{n0e}:.%

\vspace{0.5cm}

Based on these calculations, it is interesting to compare the density of superconducting carriers
$n_0$  at $T = 0$, which is described by Eq.(\ref{n-0}), with the density of normal carriers $n_e(T_c)$, which are
evaporated on levels above $\mathcal{E}_F$ at $T=T_c$ and are described by Eq.(\ref{net}).
\bigskip

\begin{figure}
\hspace{1.5cm}
\includegraphics[scale=0.5]{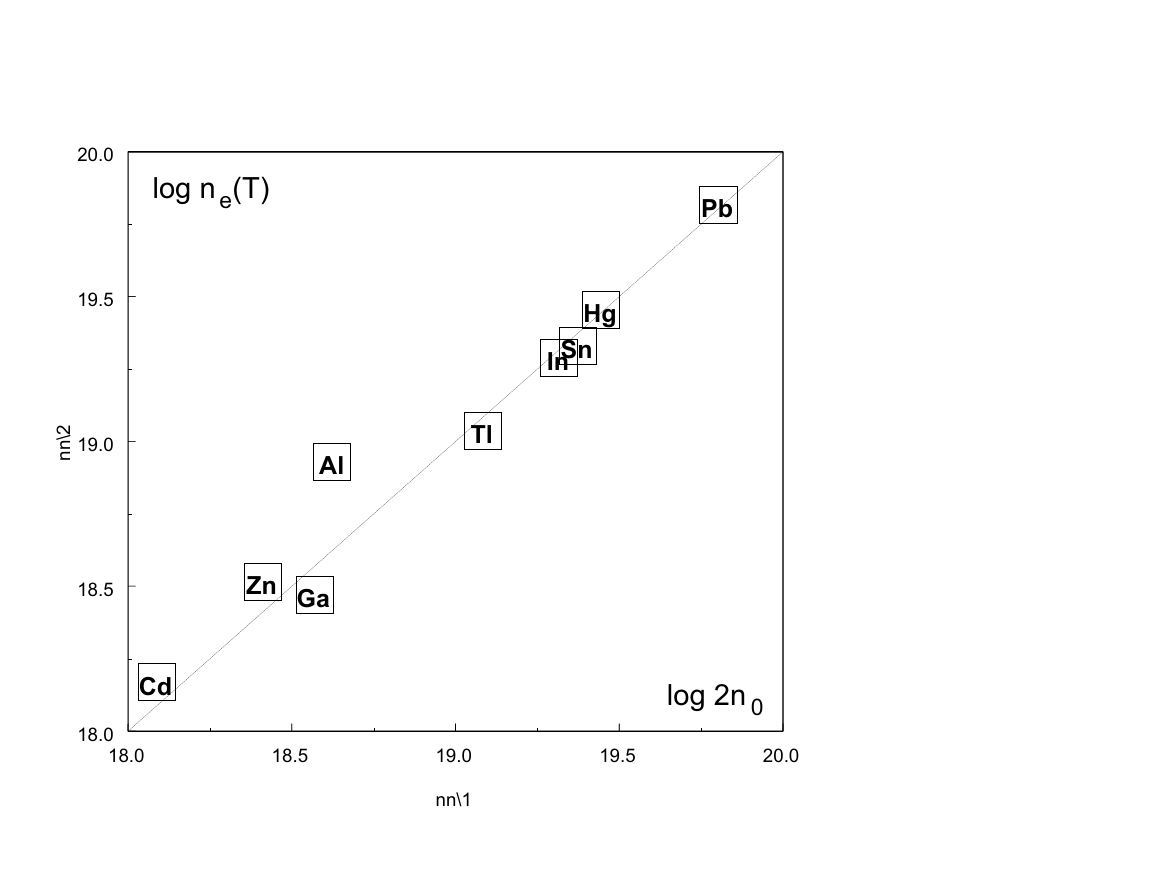}
\caption {The comparison of the number of superconducting carriers at $T=0$ with the number of thermally activated electrons at $T=T_c$.}\label{n0Ne}
\end{figure}

\bigskip

This comparison is shown in Table ({\ref{nna}}) and Fig.{\ref{n0Ne}}.
 (Data has been taken from the tables \cite{Ketterson},\cite{Pool}).

\bigskip
\begin{table}
\centering
\begin{tabular}{||c|c|c|c||}\hline\hline
  superconductor&$n_0$&$n_e({T_c})$&$2n_0/n_e(T_c)$\\\hline
  Cd &$6.11\cdot 10^{17}$&$1.48\cdot 10^{18}$&0.83\\
  Zn &$1.29\cdot 10^{18}$&$3.28\cdot 10^{18}$&0.78\\
  Ga &$1.85\cdot 10^{18}$&$2.96\cdot 10^{18}$&1.25\\
  Al &$2.09\cdot 10^{18}$&$8.53\cdot 10^{18}$&0.49\\
  Tl &$6.03\cdot 10^{18}$&$1.09\cdot 10^{19}$&1.10\\
  In &$1.03\cdot 10^{19}$&$1.94\cdot 10^{19}$&1.06\\
  Sn &$1.18\cdot 10^{19}$&$2.14\cdot 10^{19}$&1.10\\
  Hg &$1.39\cdot 10^{19}$&$2.86\cdot 10^{19}$&0.97\\
  Pb &$3.17\cdot 10^{19}$&$6.58\cdot 10^{19}$&0.96\\ \hline\hline
  \end{tabular}
\caption{The comparison of the superconducting carriers density at $T=0$ with the density
of thermally activated electrons at $T=T_c$}
\label{nna}
\end{table}

\bigskip

From the data described above, we can obtain the condition of destruction of superconductivity, after heating  for  superconductors of type-I, as written in the equation:
\begin{equation}
{n_e}{({T_c})}\simeq 2{n_0}\label{nn}
\end{equation}

\subsubsection[The sound velocity]{The sound velocity of the zero-point oscillations condensate}
The wavelength of zero-point oscillations $\Lambda_0$  in this model is an analogue of the Pippard coherence length in the BCS.  As usually accepted \cite{Ketterson}, the coherence length $\xi = \frac{\hbar v_F} {4\Delta_0}$. The ratio of these lengths,  taking into account Eq.(\ref{TT0}), is simply the constant:
\begin{equation}
\frac{\Lambda_0}{\xi}\approx 8\pi^2\alpha^2\approx \cdot 10^{-3}.
\end{equation}

\vspace{0.5cm}

The attractive forces arising between the dipoles located at a distance $\frac{\Lambda_0}{2}$ from each other and vibrating in opposite phase, create pressure in the system:
\begin{equation}
P\simeq\frac{d\Delta_0}{dV}\simeq \frac{d_\Omega^2}{L_0^6}.%\label{tctf}
\end{equation}
In this regard, sound into this condensation should propagate with the velocity:
\begin{equation}
c_s\simeq\sqrt{ \frac{1}{2m_e}\frac{dP}{dn_0}}.%\label{tctf}
\end{equation}
After the appropriate substitutions, the speed of sound in the condensate can be expressed through the Fermi velocity of electron gas
\begin{equation}
c_s\simeq \sqrt{2\pi^2 \alpha^3}v_F\simeq 10^{-2} v_F.%\label{tctf}
\end{equation}

The condensate particles moving with velocity $c_S$ have the kinetic energy:
\begin{equation}
{2m_e c_s^2}\simeq\Delta_0.\label{cs}
\end{equation}
Therefore, by either heating the condensate  to the critical temperature when each of its volume obtains the energy $\mathcal{E}\approx n_0\Delta_0$, or initiating the current of its particles with a velocity exceeding $c_S$, can achieve the destruction of the condensate.
(Because the condensate of charged particles oscillations is considered, destroying its coherence can be also obtained at the application of a sufficiently strong magnetic field. See below.)

\subsubsection{The relationship $\Delta_0/kT_c$}\label{Delta-TcE}
From Eq.(\ref{nn}) and taking into account Eqs.(\ref{tcc}),(\ref{net})  and (\ref{n-0}), which were obtained for condensate,  we have:
\begin{equation}
\frac{\Delta_0}{kT_c}\simeq 1.86 .
\end{equation}

 This estimation of the relationship  $\Delta_0/kT_c$ obtained for condensate has a satisfactory agreement with the measured data \cite{Pool}, for type-I superconductors as listed in Table (\ref{DeTe}).\footnote{In the BCS-theory  $\frac{\Delta_0}{kT_c}\simeq1.76$.}

\bigskip

\begin{table}
\centering
\begin{tabular}{||c|c|c|c||}\hline\hline
&&&\\%\hline
  superconductor&$T_c$,K&$\Delta_0$,mev&$\frac{\Delta_0}{kT_c}$\\%\hline
  &&&\\\hline
  Cd &0.51&0.072&1.64\\
  Zn &0.85&0.13&1.77\\
  Ga &1.09&0.169&1.80\\
  Tl &2.39&0.369&1.79\\
  In &3.41&0.541&1.84\\
  Sn &3.72&0.593&1.85\\
  Hg &4.15&0.824&2.29\\
  Pb &7.19&1.38&2.22\\ \hline\hline
\end{tabular}
\caption{The value of ratio $\Delta_0/kT_c$ obtained experimentally for type-I superconductors}
\label{DeTe}
\end{table}

%\newpage
\subsection{Another superconductors}
\subsubsection{About type-II superconductors}{The estimation of properties of type-II superconductors}

In the case of type-II superconductors the situation is more complicated.

In this case, measurements show that these metals have an electronic specific heat that has an order of value greater than those calculated on the base of free electron gas model.

 The peculiarity of these metals is associated with the specific structure of their ions.
  They are transition metals with unfilled inner d-shell (see Table \ref{mII}).

It can be assumed that the increase in the electronic specific heat of these metals should be associated with a characteristic interaction of free electrons with the electrons of the unfilled d-shell.

\bigskip

\begin{table}
\centering
\begin{tabular}{||c|c||}\hline\hline
  superconductors &electron shells\\\hline
  $Ti$ &$3d^2 ~4s^2$\\
  $V$ &$3d^3  ~4s^2$\\
  $Zr$ &$4d^2~ 5s^2$\\
  $Nb$ &$4d^3 ~5s^2$\\
  $Mo$ &$4d^4~ 5s^2$\\
  $Tc$ &$4d^5 ~5s^2$\\
  $Ru$ &$4d^6~ 5s^2$\\
  $La$ &$5d^1 ~6s^2$\\
  $Hf$ &$5d^2 ~6s^2$\\
  $Ta$ &$5d^3 ~6s^2$\\
  $W$ &$5d^4~6s^2$\\
  $Re$ &$5d^5 ~6s^2$\\
  $Os$ &$5d^6 ~6s^2$\\
  $Ir$ &$5d^7~6s^2$\\\hline\hline
\end{tabular}
\caption{The external electron shells of elementary type-II superconductors}
\label{mII}
\end{table}

\bigskip

Since the heat capacity of the ionic lattice of metals is negligible at low temperatures,
only the electronic subsystem is thermally active .

At $T = 0$ the superconducting careers populates the energetic level $\mathcal{E}_F-\Delta_0$.
During the  destruction of superconductivity through heating, an each heated career increases its thermal vibration.
If the effective velocity  of vibration is $v_t$, its kinetic energy:
\begin{equation}
\mathcal{E}_k=\frac{mv_t^2}{2}\simeq \Delta_0
\end{equation}

Only a fraction of the heat energy transferred to the metal is consumed in order to increase the kinetic energy of the electron gas in the transition metals.

Another part of the energy will be spent on the magnetic interaction of a moving electron.

At contact with the d-shell electron, a freely moving electron  induces onto it the magnetic field of the order of value:
\begin{equation}
H\approx \frac{e}{r_c^2}\frac{v}{c}.%\label{hhc}
\end{equation}
The magnetic moment of d-electron is approximately equal to the Bohr magneton.
Therefore the energy of the magnetic interaction between a moving electron of conductivity and a d-electron is approximately equal to:
\begin{equation}
\mathcal{E}_\mu \approx \frac{e^2}{2r_c}\frac{v}{c}.%\label{hhc}
\end{equation}
This energy is not connected with the process of destruction of superconductivity.

Whereas, in metals with a filled d-shell (type-I superconductors), the whole heating energy  increases the kinetic energy of  the conductivity electrons and only a small part of the heating energy is spent on it in transition metals:
\begin{equation}
\frac{\mathcal{E}_k}{\mathcal{E}_\mu+\mathcal{E}_k}\simeq \frac{m v_t}{h}a_B.\label{kmuq}
\end{equation}
So approximately
\begin{equation}
\frac{\mathcal{E}_k}{\mathcal{E}_\mu+\mathcal{E}_k}\simeq \frac{a_B}{L_0}.\label{kmu}
\end{equation}

Therefore, whereas the dependence of the gap in type-I superconductors  from  the heat capacity is defined by Eq.(\ref{tcc}), it is necessary to  take into account the relation Eq.(\ref{kmu}) in type-II superconductors for the determination of this gap dependence.
As a result of this estimation, we can obtain:
\begin{equation}
\Delta_0\simeq \Theta \gamma^2\left(\frac{\mathcal{E}_k}{\mathcal{E}_\mu+\mathcal{E}_k}\right)\simeq \Theta \gamma^2\left(\frac{a_B}{L_0}\right)\frac{1}{2},\label{delta2}
\end{equation}
where $1/2$  is the fitting parameter.

 The comparison of the results of these calculations with the measurement data (Fig.(\ref{tc2g})) shows that for the majority of type-II superconductors the estimation Eq.(\ref{delta2}) can be considered quite satisfactory.\footnote{The lowest critical temperature was measured for Mg. It is approximately equal to 1mK.  Mg-atoms in the metallic state are given two electrons into  the electron gas of conductivity. It is confirmed by the fact that the pairing of these electrons, which manifests itself in the measured value of the flux quantum \cite{Sharv}, is observed above $T_c$.
 It would seem that in view of this metallic Mg-ion must have electron shell like the Ne-atom.
 Therefore it is logical to expect that the critical temperature of Mg can be calculated by the formula for I-type  superconductors. But actually in order to get the value of $T_c\approx 1mK$,  the critical temperature of Mg should be calculated by the formula (\ref {delta2}), which is applicable to the description of metals with an unfilled inner shell. This suggests that the ionic core of magnesium metal apparently is not as simple as the completely filled Ne-shell.}

\begin{figure}
\includegraphics[scale=.5]{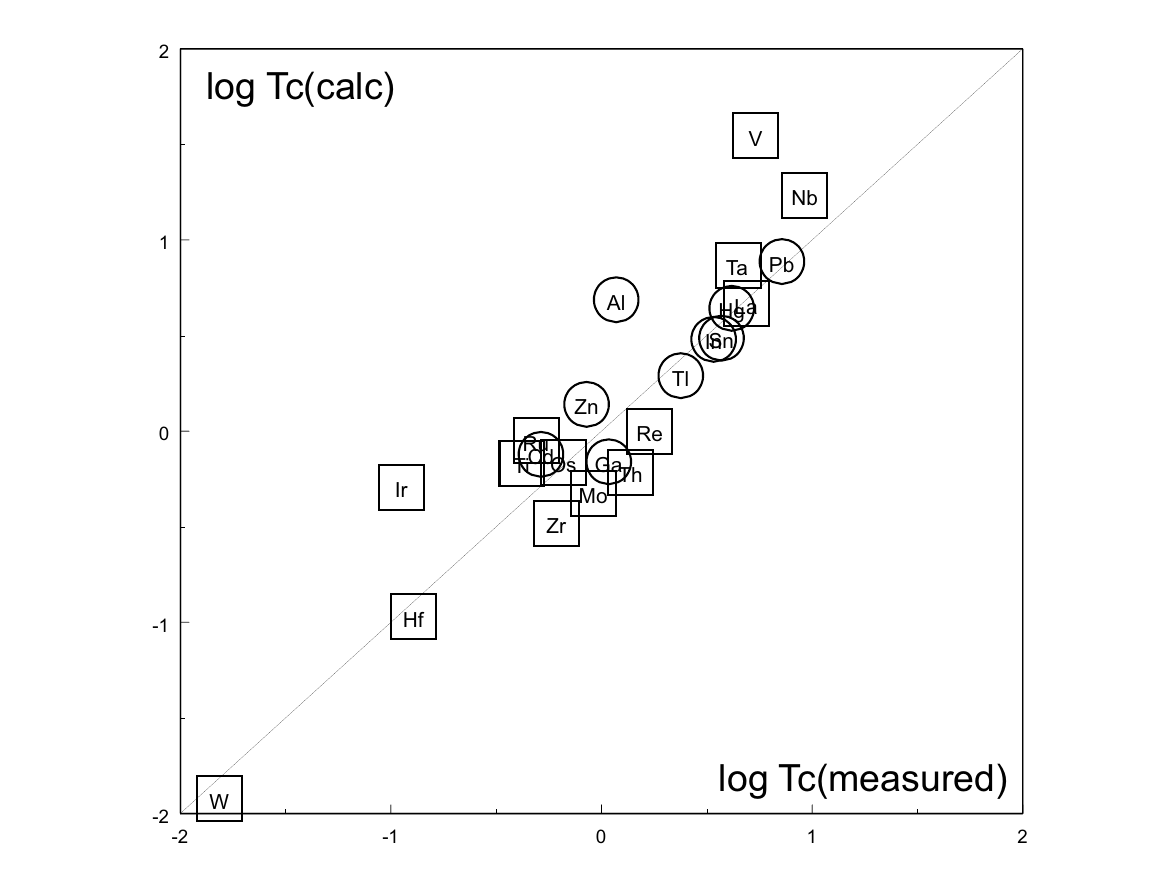}
\caption{The comparison of the calculated values of critical temperatures of superconductors with measurement data.
Circles  relate to type-I superconductors, squares show type-II superconductors.
On the abscissa, the measured values of critical temperatures are plotted, on ordinate, the calculated estimations are plotted. The calculations of critical temperatures for type-I superconductors were made by using Eq.(\ref{tcc}) and
the estimations for type-II superconductors was obtained by using Eq.(\ref{delta2}).}\label{tc2g}
\end{figure}

%\newpage

\subsubsection{Alloys and high-temperature superconductors}
In order to understand the mechanism of high temperature superconductivity, it is important to establish whether the high-$T_c$ ceramics are the I or II-type superconductors, or whether they are a special class of superconductors.

 In order to determine this, we need to look at the above established dependence of critical parameters from the electronic specific heat and also consider that the specific heat of superconductors I and II-types are differing considerably.

There are some difficulties by determining the answer this way: as we  do not precisely know the density of the electron gas in high-temperature superconductors.
However, the densities of atoms in metals do not differ too much and we can use Eq.(\ref{tcc}) for  the solution of the problem  of the I- and II-types superconductors  distinguishing.

If parameters of type-I superconductors are inserted into this equation, we obtain quite a satisfactory estimation of the critical temperature (as was done above, see Fig.\ref{tc2g}). For the type-II superconductors` values, this  assessment gives an overestimated value due to the fact that type-II superconductors' specific heat has additional term associated with the magnetization of d-electrons.

This analysis therefore, illustrates a possibility where we can divide all superconductors into two groups, as is evident from the Fig.(\ref{gamma2}).

\begin{figure}
\hspace{1.5cm}
\includegraphics[scale=.4]{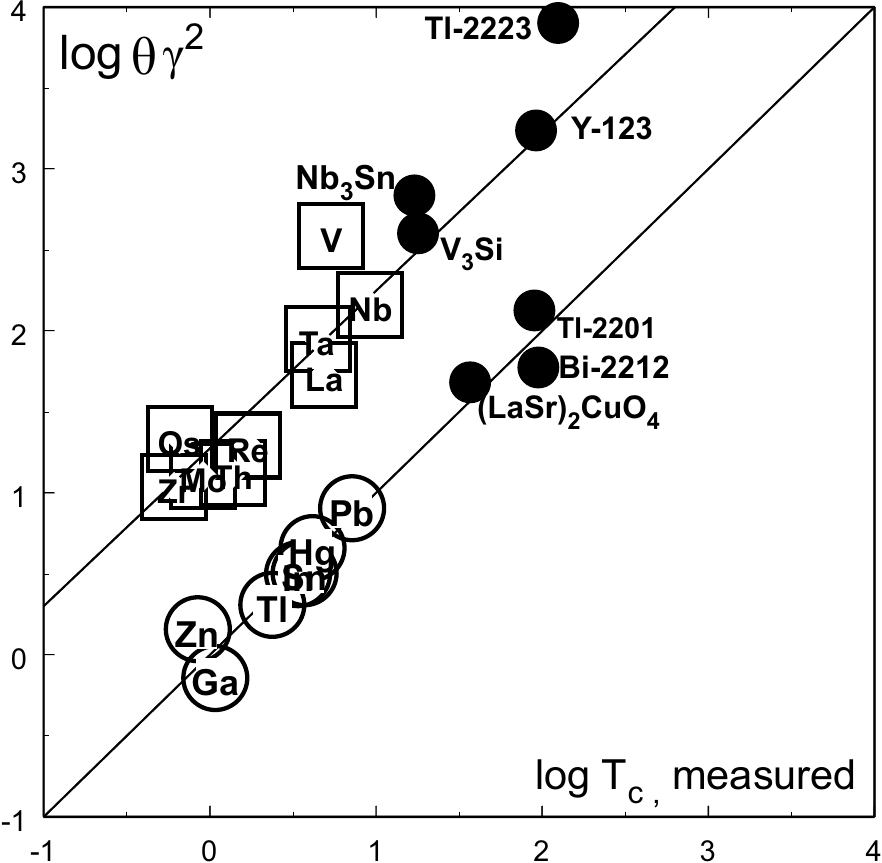}
\caption {The comparison of the calculated parameter $\Theta\gamma^2$ with the measurement of the critical temperatures of elementary superconductors and some superconducting compounds.}\label{gamma2}
\end{figure}

It is generally assumed that we  consider alloys $Nb_3Sn$ and $V_3Si$ as the type-II superconductors. This assumption seems quite normal because they are placed in close surroundings of Nb.
Some excess of the calculated critical temperature over the experimentally measured value for ceramics $Ta_2Ba_2Ca_2Cu_3O_{10}$ can be attributed to  the measured heat capacity that may have been created by not only conductive electrons, but also non-superconducting elements (layers) of ceramics. It is already known that it, as well as ceramics $YBa_2Cu_3O_7$, belongs to the type-II superconductors. However, ceramics (LaSr)$_2$Cu$_4$, Bi-2212 and Tl-2201, according to this figure should be regarded as type-I superconductors, which is unusual.

%\newpage

\subsection{About the London penetration depth}
\label{LondonE}

\subsubsection{The traditional approach to calculation of the London penetration depth}

The consideration of the London penetration depth is commonly accepted
 (see, for example \cite{Ketterson}) in several steps:

 \vspace {0.2cm}

 \underline{Step 1.}

Firstly, the action of an external electric field on free electrons is considered.
In accordance with Newton's law, free electrons gain acceleration in an electric field $\mathbf{E}$:
\begin{equation}
\mathbf{a}=\frac{e{\mathbf{E}}}{m_e}.\label{aE}
\end{equation}
The directional movement of the "superconducting" \ electron gas with the density
  $n_s$  creates the current with the density:
\begin{equation}
\mathbf{j}=en_s\mathbf{v},
\end{equation}
where $\mathbf{v}$ is the carriers velocity.

After differentiating the time and substituting this in Eq.(\ref{aE}), we obtain the first London's equation:
\begin{equation}
\frac{d}{dt}\mathbf{j}=en_s\mathbf{a}=\frac{n_s e^2}{m_e}\mathbf{E}.\label{Lo1}
\end{equation}

\underline{Step 2}.

 After application of operations $rot$ to both sides of this equation and by
 using  the Faraday's law of electromagnetic induction $rot\mathbf{E}=-\frac{1}{c}\frac{d\mathbf{B}}{dt}$
 we acquire the relationship between the current density and magnetic field:
\begin{equation}
\frac{d}{dt}\left(rot~\mathbf{j}+\frac{n_s e^2}{m_e c}\mathbf{B}\right)=0.\label{L11}
\end{equation}

\underline{Step 3.}
 By selecting the stationary solution of  Eq.(\ref{L11})
 \begin{equation}
rot~\mathbf{j}+\frac{n_s e^2}{m_e c}\mathbf{B}=0,\label{L12}
\end{equation}
and  after some simple transformations, we can conclude
 that there is a so-called London penetration depth of the magnetic field in a superconductor:

\begin{equation}
\Lambda_L=\sqrt{\frac{m_e c^2}{4\pi e^2 n_s}}.\label{lamb}
\end{equation}

\vspace{0.5cm}

\subsubsection{The London penetration depth and the density of superconducting carriers}

One of the measurable characteristics of superconductors is the London penetration depth, and for many of these superconductors it usually equals to a few hundred Angstroms \cite{Linton}. In the Table (\ref{L1T}) the measured values of $\lambda_L$ are given in the second column.

\vspace{0.5cm}

{%\scriptsize
{
%\begin{table}
\hspace{-0.5cm}\begin{tabular}{||c|c|c|c|c||} \hline\hline
                 & $\lambda_L$,$10^{-6}$cm &      $n_{s}$       &    $n_e$          & \\
  super-& measured                &    according to     &  in accordance  &$n_s/n_e$\\
        conductors & \cite{Linton}           &     Eq.(\ref{lamb}) & with Eq.(\ref{lF}) &\\\hline
  Tl             &   9.2                  &       $3.3\cdot 10^{21}$  &$1.4\cdot 10^{23}$ & 0.023    \\
  In             &   6.4                   &      $6.9\cdot 10^{21}$  &$3.0\cdot 10^{23}$ & 0.024     \\
  Sn             &   5.1                   &      $1.1\cdot 10^{22}$  &$3.0\cdot 10^{23}$ & 0.037     \\
  Hg             &   4.2                   &      $1.6\cdot 10^{22}$  &$4.5\cdot 10^{22}$ & 0.035 \\
  Pb             &   3.9                   &      $1.9\cdot 10^{22}$  &$1.0\cdot 10^{24}$ & 0.019     \\\hline\hline
\end{tabular}
%\end{table}
\label{L1T}

{Table({\ref{L1T}})}}}

\vspace{0.5cm}

If we are to use  this experimental data to calculate the density of superconducting carriers $n_s$
in accordance with the Eq.(\ref{lamb}), the results be about two orders of magnitude larger
(see the middle column of Tab.(\ref{L1T}).

Only a small fraction of these free electrons can combine into the  pairs. This is
only applicable to the electrons that energies lie  with in the thin strip of the energy spectrum near $\mathcal{E}_F$.
We can therefore expect that the concentration of superconducting carriers among all free electrons of the metal
should be at the level $\frac{n_s}{n_e}\approx 10^{-5}$ (see Eq.(\ref{n0-ne})).
These concentrations, if calculated from Eq.(\ref{lamb}), are seen to be about two orders of magnitude higher (see last column of the Table (\ref{L1T}).

Apparently, the reason for this discrepancy is because of the use of a nonequivalent transformation.
At the first stage in Eq.(\ref{aE}), the straight-line acceleration in a static electric field is considered.
If this moves, there will be no current circulation. Therefore, the application  of the operation rot in Eq.(\ref{L11}) in this case is not correct.
It does not lead to the Eq.(\ref{L12}):
\begin{equation}
\frac{rot~\mathbf{j}}{\frac{n_s e^2}{m_e c}\mathbf{B}}=-1,
\end{equation}
but instead, leads to a pair of equations:
\begin{equation}
\begin{array}{l}
rot~\mathbf{j}=0\\
\frac{n_s e^2}{m_e c}\mathbf{B}=0
\end{array}
\label{L13}
\end{equation}
and to the uncertainty:
\begin{equation}
\frac{rot~\mathbf{j}}{\frac{n_s e^2}{m_e c}\mathbf{B}}=\frac{0}{0}.
\end{equation}

\subsubsection{The magnetic energy of a moving electron}
To avoid these incorrect results, let us consider a balance of magnetic energy in a superconductor within magnetic field.
This magnetic energy is composed of energy from a penetrating external magnetic field and magnetic energy of  moving electrons.

 %\bf{The magnetic energy of a moving electron}

By using formulas \cite{Bek}, let us estimate the ratio of the magnetic and kinetic energy of an electron (the charge of $e$ and the mass $m_e$) when it moves rectilinearly with a velocity $v\ll c$.

The density of the electromagnetic field momentum  is expressed by the equation:
\begin{equation}
\mathbf{g}=\frac{1}{4\pi c}[\mathbf{E}\mathbf{H}]%\label{Be}
\end{equation}

While moving with a velocity $\mathbf{v}$, the electric charge carrying the electric field with intensity $E$ creates a magnetic field
\begin{equation}
\mathbf{H}=\frac{1}{c}[\mathbf{E}\mathbf{v}]\label{Hvc}
\end{equation}
with the density of the electromagnetic field momentum (at $v\ll c$)
\begin{equation}
\mathbf{g}=\frac{1}{4\pi c^2}[\mathbf{E}[\mathbf{v}\mathbf{E}]]=\frac{1}{4\pi c^2}\left(\mathbf{v}E^2-\mathbf{E}(\mathbf{v}\cdot \mathbf{E})\right)%\label{Be}
\end{equation}
As a result, the momentum of the electromagnetic field of a moving electron
\begin{equation}
\mathbf{G}=\int_V \mathbf{g}dV=
\frac{1}{4\pi c^2}\left(\mathbf{v}\int_V E^2 ~dV - \int_V \mathbf{E}~E~v~ cos\vartheta ~dV \right)\label{cos}
\end{equation}
The integrals are taken over the entire space, which is occupied by particle fields, and $\vartheta$ is the angle between the particle velocity and the radius vector of the observation point. By calculating the last integral in the condition of the axial symmetry with respect to $\mathbf{v}$, the contributions from the components of the vector $\mathbf{E}$, which is perpendicular to the velocity, cancel each other for all pairs of elements of the space (if they located diametrically opposite on the magnetic force line). Therefore, according to  Eq.(\ref{cos}), the component of the field which is collinear to $\mathbf{v}$
\begin{equation}
\frac{E~cos\vartheta \cdot \mathbf{v}}{v}%\label{Be}
\end{equation}
can be taken instead of the vector $\mathbf{E}$.
By taking this information into account, going over to the spherical coordinates and integrating over angles, we can obtain
\begin{equation}
\mathbf{G}=
\frac{\mathbf{v}}{4\pi c^2}\int_{r}^\infty E^2\cdot 4\pi r^2~dr%\label{cos}
\end{equation}
If we limit the integration of the field by the Compton electron radius $r_C=\frac{\hbar}{m_e c}$, ~ \footnote{Such effects as the pair generation force us to consider the radius of the "quantum electron" \ as approximately equal to Compton radius \cite{Messia}.} then $v\ll c$, and we obtain:
\begin{equation}
\mathbf{G}=\frac{\mathbf{v}}{4\pi c^2}\int_{r_C}^\infty E^2\cdot 4\pi r^2~dr=
\frac{\mathbf{v}}{c^2} \frac{e^2}{r_C}.%\label{cos},
\end{equation}
In this case by taking into account Eq.(\ref{Hvc}), the magnetic energy of a slowly moving electron pair is equal to:
\begin{equation}
\mathcal{E}=\frac{{v}{G}}{2}=
\frac{{v^2}}{c^2} \frac{e^2}{2r_C}=\alpha\frac{m_e v^2}{2}.\label{EEalfa}
\end{equation}

\subsubsection{The magnetic energy and the London penetration depth}
\label{London-new2E}
%\bf{The magnetic energy and the London penetration depth}
The energy of external magnetic field into volume $dv$:
\begin{equation}
\mathcal{E}=\frac{H^2}{8\pi}dv.%\label{L12}
\end{equation}
At density of superconducting carriers $n_s$, their magnetic energy per unit volume in accordance with (\ref {EEalfa}):
\begin{equation}
\mathcal{E}_H\simeq\alpha n_s\frac{m_2 v^2}{2}=\alpha\frac{m_e j_s^2}{2n_s e},%\label{L12}
\end{equation}
where $j_s=2e n_s v_s$ is the density  of a current of superconducting carriers.

Taking into account the Maxwell equation
\begin{equation}
\mathbf{rot H}=\frac{4\pi}{c}\mathbf{j}_s,%\label{L12}
\end{equation}
the magnetic energy of moving carriers can be written as
\begin{equation}
\mathcal{E}_H\simeq \frac{\widetilde{\Lambda}^2}{8\pi}(rot H)^2,%\label{L12}
\end{equation}
where we introduce the notation
\begin{equation}
\widetilde{\Lambda}=\sqrt{\alpha \frac{m_e c^2}{4\pi n_s e^2}}=\sqrt{\alpha}\Lambda_L.\label{lam}
\end{equation}
In this case, part of the free energy of the superconductor connected with the application of a magnetic field is equal to:
\begin{equation}
\mathcal{F}_H=\frac{1}{8\pi}\int_V\left(H^2+\widetilde{\Lambda}^2(rot H)^2\right)dv.%\label{L12}
\end{equation}
At the minimization of the free energy, after some simple transformations we obtain
\begin{equation}
\mathbf{H}+\widetilde{\Lambda}^2 \mathbf{rot rot H}=0,%\label{L12}
\end{equation}
thus  $\widetilde{\Lambda}$ is the depth of magnetic field penetration into the superconductor.

In view of Eq.(\ref{n-0}) from Eq.(\ref{lam}) we can estimate the values of London penetration depth (see table (\ref{London2})).
The consent of the obtained values with the measurement data can be considered quite satisfactory.

\vspace{0.5cm}

\begin{table}
\centering
\begin{tabular}{||c|c|c|c||} \hline\hline
&&&\\
super-&$\lambda_L$,$10^{-6}$cm &$\widetilde{\Lambda}$,$10^{-6}$cm&\\
conductors     & measured \cite{Linton} & calculated&$\widetilde{\Lambda}/\lambda_L$\\
 &                         &Eq.(\ref{lam})&\\\hline
  Tl  &    9.2   &     11.0&1.2 \\
  In  &    6.4   &     8.4&1.3 \\
  Sn  &    5.1   &     7.9&1.5 \\
  Hg  &    4.2   &     7.2&1.7 \\
  Pb  &    3.9   &     4.8&1.2 \\\hline\hline
\end{tabular}
\caption{Corrected values of London penetration depth}
\label{London2}
\end{table}

\vspace{0.5cm}

The resulting refinement can be important for estimations within the frame of Ginzburg-Landau theory, where the London
 penetration depth is used as a comparison of calculations and specific parameters of superconductors.

%\newpage

\vspace{0.5cm}
\newpage
\subsection{Three words to experimenters}

\subsubsection[Is the room-temperature superconductivity possible?]{Why creation of room-temperature superconductors is very problematic?}
%\bf
The understanding of the mechanism of the superconducting state should open a way towards finding a solution to the technological problem. This problem was just a dream in the last century:
 the dream to create a superconductor that would be easily  produced (in the sense of ductility) and had  high critical temperature.

In order to move towards  this goal,
it is important firstly to understand the mechanism that limits the  critical properties of superconductors.

Let us consider a superconductor with a large limiting current.
The length of their localisation determines the limiting  momentum of  superconducting carriers:
\begin{equation}
p_c\simeq\frac{2\pi \hbar}{L_0}%\label{L12}
\end{equation}
Therefore, by using  Eq.(\ref{cs}), we can compare the critical velocity of superconducting carriers with the sound velocity:
\begin{equation}
v_c=\frac{p_c}{2m_e}\simeq c_s%.\label{f0}
\end{equation}
 and these velocities are both about a hundred times smaller than the Fermi velocity.

The sound velocity in the crystal lattice of  metal $v_{s}$, in accordance with the Bohm-Staver relation \cite{Ashkr}, has  approximately the same value:
\begin{equation}
v_s\simeq \frac{k T_D}{E_F}v_F\simeq 10^{-2}{v_F}.%\label{f0}
\end{equation}
This therefore makes it possible to consider superconductivity being destroyed as a superconducting carrier overcomes
the sound barrier.
That is, if they moved without friction at a speed that was less than the speed of the sound, after they gained speed and the speed of sound was surpassed,
 then they acquire a mechanism of friction.

 Therefore, it is conceivable that if the speed of sound in the metal lattice $v_s<c_s$, then it would create a restriction on the limiting current in superconductor.

 If this is correct, then superconductors with high critical parameters should have not only a high Fermi energy of their electron gas, but also a high speed of sound in their lattice.

It is in agreement with the fact that  ceramics have higher elastic moduli compared to metals and alloys, and also posses  much higher critical temperatures (Fig.{\ref{vsg}}).

\begin{figure}
\hspace{-0.5cm}
\includegraphics[scale=0.4]{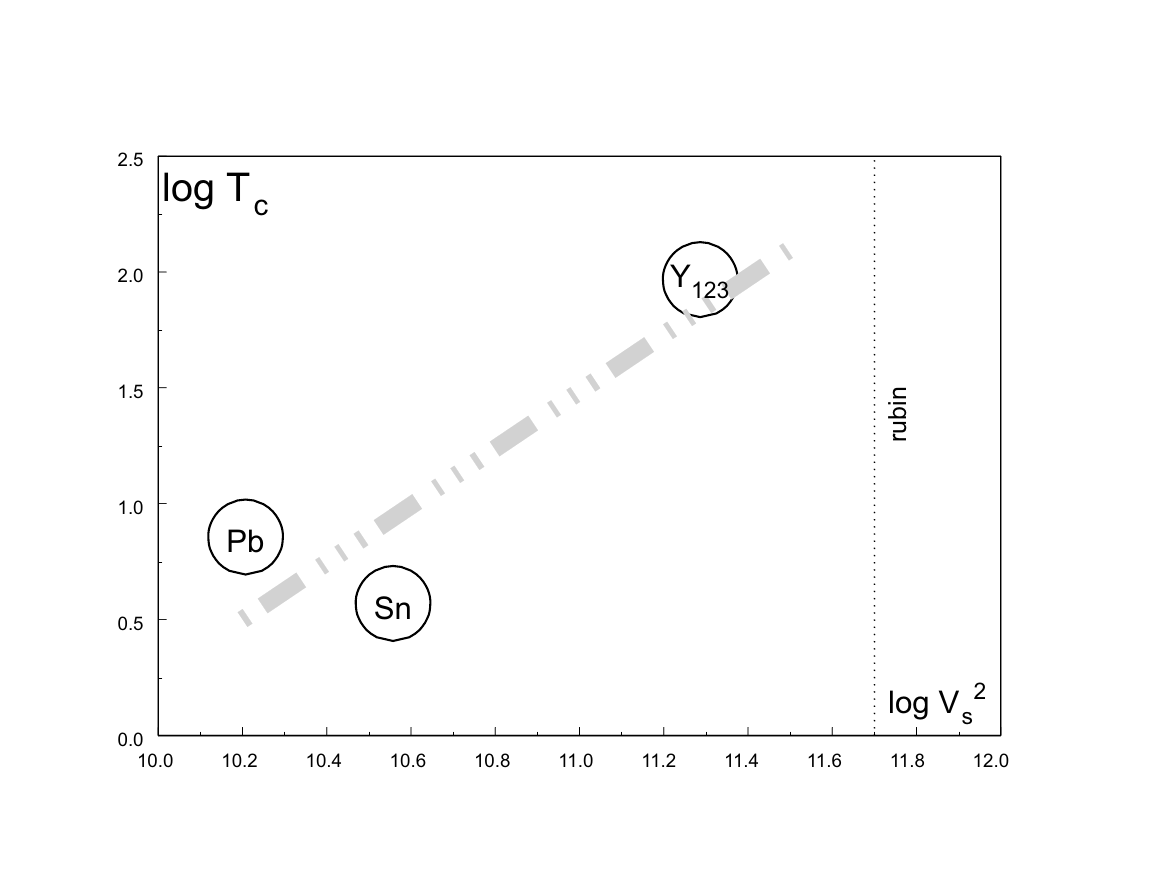}
\caption{The schematic representation of the dependence of critical temperature on the speed of sound in superconductors. On the ordinate, the logarithm of the superconductor's critical temperature is shown. On the abscissa, the logarithm of the square of the speed of sound  is shown (for Sn and Pb - the transverse velocity of sound is shown, because it is smaller).  The speed of sound in a film   was used for yttrium-123 ceramics. The dashed line shows the value of the transverse velocity of sound in sapphire, as some estimation of the limit of its value.
It  can be seen that this estimation leads to the restriction on the critical temperature in the range about $0^o$C - the dot-dashed line.}\label{vsg}
\end{figure}

\vspace{1cm}

The dependence of the critical temperature on the square of the speed of sound \cite{Gol} is illustrated in Fig.(\ref{vsg}).

  This figure, which can be viewed only as a rough estimation due to the lack of necessary experimental data, shows that the elastic modulus of ceramics with a critical temperature close to the room temperature, should be close to the elastic modulus of sapphire, which is very difficult to achieve.

In addition, such ceramics would be deprived from another important quality: their adaptability.
Indeed, in order to obtain a thin wire, we  require a plastic superconductor.

A solution of this problem would be to find  a material that possesses an acceptably high critical temperature (above 80K) and  experiences a phase transition at even higher temperature of heat treatment.
It would be possible to make a thin wire from a superconductor near the point of phase transition, as the elastic modules  are typically  not very strong at this stage.

\subsubsection{Magnetic electron pairing}
The described formation of mechanism for the superconducting state provides a possibility to obtain an estimations  of the critical parameters of superconductors, which in most cases is in satisfactory agreement with measured data. For some superconductors, this agreement is stronger, and for other, such as Ir, Al, V (see Fig.(\ref{tc2g})), it is expedient to carry out further theoretical and experimental studies due to causes of deviations.

The mechanism of magnetic electron pairing is also very important for further clarification of this phenomenon.

As it was found earlier, in the cylinders made from  certain superconducting metals (Al\cite{Shab} and Mg\cite {Sharv}), the observed magnetic flux quantization has exactly the same period above $T_c$ and that below $T_c$. The authors of these studies attributed this to the influence of a special effect. It seems more natural to assume  that the stability of the  period is a result of the pairing of electrons due to magnetic dipole-dipole interaction continuing to exist at temperatures above $T_c$, despite the disappearance of the material's superconducting properties. At this temperature,  the coherence of the zero-point fluctuations is destroyed, and with it the superconductivity is destroyed.

The pairing of electrons due to dipole-dipole interaction should  be absent in the monovalent metals. In these metals, the   conduction electrons are localized in the lattice at very large distances from each other.

It is therefore interesting to compare the  period of quantization in these two cases.
In a thin cylinder made from a superconductor, such as Mg, the quantization period above $T_c$ is equal to $\frac{2\pi\hbar c}{2e}$. In the same cylinder of a noble metal (such as gold), the sampling period should be twice as large.

\subsubsection[The effect of isotopic substitution]{The effect of isotopic substitution on the condensation of zero-point oscillations}
The attention of experimentalists could be attracted
 to the isotope effect in superconductors, which served as a starting point of the B-BCS theory.
 In the 1950s, it had been experimentally established that there was a dependence of the critical temperature of superconductors due to the mass of the isotope. As the effect depends on the ionic mass, this is considered to be due to  the fact that it is based on the vibrational (phonon) process.

The isotopic effect for a number of type-I superconductors such as $Zn, Sn, In, Hg, Pb $, can be described by the relationship:
\begin{equation}
\sqrt{M_i}T_c=const,\label{ise}
\end{equation}
where $M_i$ is the mass of the isotope, $T_c$ is the critical temperature.
The isotope effect in other superconductors can either be described by other dependencies, or be totally absent.

In recent decades, however, the effects associated with the replacement of isotopes in the metal lattice have been studied in detail.
It was shown that for many metals the zero-point oscillations of  ions in the lattice are non-harmonical.
Therefore, the isotopic substitution can directly affect the lattice parameters, the density of the lattice and the density of the electron gas in the metal, on its Fermi energy  and on other  properties of the electronic subsystem.

The direct study of the effect of isotopic substitution on the lattice parameters of superconducting metals has not been carried out.

The results of measurements made on $Ge$, $Si$, diamond and light metals, such as $Li$ \cite{Kogan}, \cite{Inyu} (researchers prefer to study crystals, where the isotope effects are large, and it is easier to carry out appropriate measurements), show that there is square-root dependence of the force constants on the isotope mass, which was required by Eq.(\ref{ise}). The same dependence of the force constants on the mass of the isotope has been found in tin \cite{Wang}.

Unfortunately, no direct experiments of the effect of isotopic substitution on the electronic properties (such as the electronic specific heat and the Fermi energy), exist for metals substantial for our consideration.

Let us consider what should be expected in such measurements. A convenient choice for the superconductor is mercury, as it has many isotopes and their isotopic effect has been carefully measured back in the 1950s  as aforementioned.

The linear dependence of the critical temperature of a superconductor on its Fermi energy (Eq.(\ref{TcE})) and the existence  of the isotopic effect suggests the dependence of the density in the  crystal lattice from the mass of the isotope.

Even then, it was found that the isotopic effect is described by Eq.(\ref{ise}) in only a few superconductors. In others, it displays different values, and therefore in a general case it can be described by introducing  the parameter
$\mathfrak{a}$:
\begin{equation}
M_i^\mathfrak{a} T_c=const
\end{equation}
At taking into account Eq.(\ref{TcE}), we can write
\begin{equation}
T_c\sim \mathcal{E}_F\sim n_e^{2/3}
\end{equation}

The parameter $l$ which characterizes the ion lattice obtains an increment
$\Delta l$ with an isotope substitution:
\begin{equation}
\frac{\Delta l}{l} = -\frac{\mathfrak{a}}{2}\cdot \frac{\Delta M_i}{M_i},
\end{equation}
where ${M_i}$ and $\Delta M_i$ are the mass of isotope and its increment.

It is generally accepted that in an accordance with the terms of the phonon mechanism, the parameter $\mathfrak{a}\approx \frac{1}{2}$ for mercury.
 However, the analysis of experimental data \cite{Maxwell}-\cite{Serin} (see Fig.(\ref{Hgg})) shows that this parameter is actually closer to $1/3$. Accordingly, one can expect that the  ratio of the mercury parameters is close to:
\begin{equation}
\frac{\left(\frac{\Delta l}{l}\right)}{\left(\frac{\Delta M_i}{ M_i}\right)}\approx -\frac{1}{6} .%\label{}
\end{equation}

\begin{figure}
\hspace{1cm}
\includegraphics[scale=0.5]{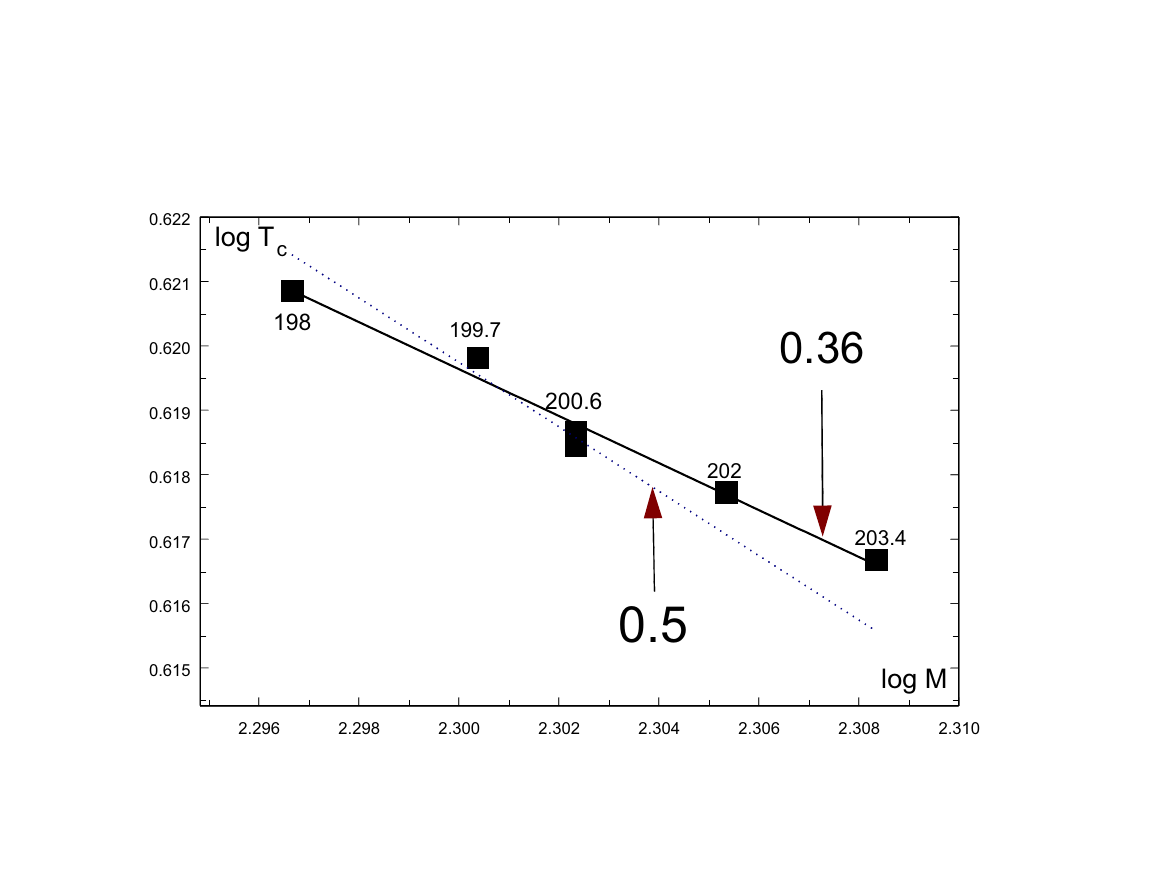}
\caption {The isotope effect in mercury.
The solid line is obtained by the sparse-squares technique.
In accordance with the phonon mechanism, the coefficient  $\mathfrak{a}$ must be about 1/2 (the dotted line).
As it can be seen, this coefficient is in reality  approximately equal to  1/3.}\label{Hgg}
\end{figure}

\newpage

\section[Superfluidity and zero-point oscillations]{Superfluidity as a sequence of  ordering of zero-point oscillations}
\label{HeE}
\subsection{Zero-point oscillation of He atoms and superfluidity}

%\bf{}
The main features of superfluidity of liquid helium became clear few decades ago
\cite{Landau}, \cite{Halat}. L.D.Landau explains this phenomenon as the manifestation of a quantum behavior of the macroscopic object.

However, the causes and mechanism of the formation of superfluidity  are not clear till our days.
There is no explanation why the $\lambda$-transition in helium-4 occurs at about 2 K, that is  about twice less than   its boiling point:
\begin{equation}
\frac{T_{boiling}}{T_\lambda}\approx 1.94,\label{f022}
\end{equation}
while for helium-3, this transition is observed only at temperatures  about a thousand times smaller.

 The related phenomenon, superconductivity, can be regarded as superfluidity of a charged liquid. It can be quantitatively described considering it as the consequence of ordering of zero-point oscillations of electron gas. Therefore it seems appropriate to consider superfluidity from the same point of view.

Atoms in liquid helium-4 are electrically neutral, as they have no dipole moments and do not form molecules.
  Yet some electromagnetic mechanism should be responsible for phase transformations of liquid helium (as well as in other condensed substance where phase transformations are related to the changes of energy of the same scale).

 F. London has demonstrated already  in the 1930's \cite{FLondon}, that there is an interaction between  atoms in the ground state, and this interaction is of a quantum nature. It can be considered as a kind of  the Van-der-Waals interaction.
 Atoms in their ground state (T = 0) perform zero-point oscillations. F.London was considering  vibrating atoms as three-dimensional oscillating dipoles which  are connected to each other by the electromagnetic interaction. He proposed the name the dispersion interaction for this interaction of atoms in the ground state.

\subsection{The dispersion effect in interaction of atoms in the ground state}
Following F.London \cite{FLondon}, let us consider two spherically symmetric atoms without non-zero average dipole moments.  Let us suppose that at some time the charges of these atoms are  fluctuationally displaced from the equilibrium states:
\begin{equation}
r_1=(x_1,y_1,z_1)\nonumber
\end{equation}
\begin{equation}
r_{2}=(x_{2},y_{2},z_{2})\nonumber
\end{equation}
If atoms are located along the Z-axis at the distance $L$ of each other, their potential energy can be written as:
\begin{equation}
\mathcal{H}=\underbrace{\frac{e^2 r_1^2}{2a}+\frac{e^2 r_{2}^2}{2a}}_{elastic~
dipoles~energy}
+\underbrace{\frac{e^2}{L^3}(x_1 x_{2}+y_1y_{2}-2z_1z_{2})}_{elastic~dipoles~interaction}.\label{q0}
\end{equation}
where $a$ is the atom polarizability.

The Hamiltonian can be diagonalized by using the normal coordinates of symmetric and antisymmetric displacements:
\begin{displaymath}
r_s\equiv
\left\{
\begin{array}{ll}
x_s=\frac{1}{\sqrt{2}}(x_1+x_2) \\
y_s=\frac{1}{\sqrt{2}}(y_1+y_2) \\
z_s=\frac{1}{\sqrt{2}}(z_1+z_2) \\
\end{array}
\right.
\end{displaymath}
and
\begin{displaymath}
r_a\equiv
\left\{
\begin{array}{ll}
x_a=\frac{1}{\sqrt{2}}(x_1-x_2) \\
y_a=\frac{1}{\sqrt{2}}(y_1-y_2) \\
z_a=\frac{1}{\sqrt{2}}(z_1-z_2) \\
\end{array}
\right.
\end{displaymath}
This yields
\begin{displaymath}
\begin{array}{ll}
x_1=\frac{1}{\sqrt{2}}(x_s+x_a) \\
y_1=\frac{1}{\sqrt{2}}(y_s+y_a) \\
z_1=\frac{1}{\sqrt{2}}(z_s+z_a) \\
\end{array}
\end{displaymath}
and
\begin{displaymath}
\begin{array}{ll}
x_2=\frac{1}{\sqrt{2}}(x_s-x_a) \\
y_2=\frac{1}{\sqrt{2}}(y_s-y_a) \\
z_2=\frac{1}{\sqrt{2}}(z_s-z_a) \\
\end{array}
\end{displaymath}
As the result of this change of variables we obtain:
{
%\scriptsize
\begin{eqnarray}
\mathcal{H}=\frac{e^2}{2a}(r_s^2+r_a^2)+
\frac{e^2}{2L^3}(x_s^2+y_s^3-2z_s^2-x_a^2-y_a^2+2z_a^2)= \nonumber\\
=\frac{e^2}{2a}\left[\left(1+\frac{a}{L^3}\right)(x_s^2+y_s^2)+ % \nonumber
\left(1-\frac{a}{L^3}\right)(x_a^2+y_a^2)+ \right. \\ %\nonumber
+\left(1-2\frac{a}{L^3}\right)z_s^2 +
\left.\left(1+2\frac{a}{L^3}\right)z_a^2\right].\nonumber\label{q01}
\end{eqnarray}}
Consequently, frequencies of oscillators depend on their  orientation and they are determined by the equations:
{
%\scriptsize
\begin{eqnarray}
\Omega_{0x}^{s\atop a}=\Omega_{0y}^{s\atop a}=\Omega_0\sqrt{1\pm\frac{a}{L^3}}\approx\Omega_0\left({1\pm\frac{a}{L^3}}-\frac{a^2}{8L^6}\pm...\right)  \\
\Omega_{0z}^{s\atop a}=\Omega_0\sqrt{1\mp\frac{2a}{L^3}}\approx\Omega_0\left({1\mp\frac{a}{L^3}}-\frac{a^2}{2L^6}\mp...\right),
\end{eqnarray}}
where
\begin{equation}
\Omega_0=\frac{2\pi e}{\sqrt{ma}}%\nonumber
\end{equation}
is natural frequency of the electronic shell of the atom (at $L\rightarrow\infty$).
The energy of zero-point oscillations is
\begin{equation}
\mathcal{E}=\frac{1}{2}\hbar(\Omega_0^s+\Omega_0^a)%\label{EE}
\end{equation}
It is easy to see that the description of interactions between neutral atoms do not contain terms $\frac{1}{L^3}$, which are characteristics for the interaction of zero-point oscillations in the electron gas (Eq.(\ref{Lz3})) and
which are responsible for the occurrence of superconductivity. \\
The terms that are proportional to $\frac{1}{L^6}$ manifest themselves in interactions of neutral atoms.\\

It is important to emphasize that the energies of interaction are different for  different orientations of zero-point oscillations. So the interaction of zero-point oscillations oriented along the direction connecting the atoms leads to their attraction with energy:
\begin{equation}
\mathcal{E}_z = - \frac{1}{2}\hbar\Omega_0\frac{A^2}{L^6},\label{f1}
\end{equation}
while the summary energy of the attraction of the oscillators of the perpendicular directions (x and y)  is equal to one half of it:
\begin{equation}
\mathcal{E}_{x+y} = - \frac{1}{4}\hbar\Omega_0\frac{A^2}{L^6}\label{f2}
\end{equation}
(the minus sign is taken here because  for this case   the opposite direction of dipoles  is energetically favorable).

\vspace{1cm}

\subsection[The main characteristics of  superfluid helium]{The estimation of main characteristic parameters of  superfluid helium}
\subsubsection{The main characteristic parameters of the zero-point oscillations of atoms in superfluid helium-4}

There is no repulsion in a gas of neutral bosons.
Therefore, due to attraction between the atoms at temperatures below
\begin{equation}
T_{boil}=\frac{2}{3k}\mathcal{E}_z%\label{f1}
\end{equation}
this gas collapses and a liquid  forms.

At twice lower temperature
\begin{equation}
T_\lambda=\frac{2}{3k}\mathcal{E}_{x+y}%\label{f1}
\end{equation}
all zero-point oscillations become ordered. It creates an additional attraction and forms a single quantum ensemble.

A density of the boson condensate  is limited by  zero-point oscillations  of its atoms.
At condensation the distances between the atoms become approximately equal  to  amplitudes of zero-point oscillations.\\

Coming from it, we can calculate the basic properties of an ensemble of atoms with ordered zero-point oscillations, and compare them with measurement properties of superfluid helium.

We can assume that the radius of a helium atom is equal to the Bohr radius $a_B$, as it follows from quantum-mechanical calculations.
Therefore, the energy of electrons on the s-shell of this atom can be considered to be equal:
\begin{equation}
\hbar\Omega_0 = \frac{4e^2}{a_B}%\label{f1}
\end{equation}
As the polarizability of atom is approximately equal to its volume \cite{Fr}
\begin{equation}
A\simeq a_B^3,\label{vA}
\end{equation}
the potential energy of dispersive interaction (\ref{f2}), which causes the ordering zero-point oscillations in the ensemble of atoms, we can represent by the equation:
\begin{equation}
\mathcal{E}_{x+y} = - \frac{e^2}{a_B}a_B^6 n^2,\label{W}
\end{equation}
where the density of helium atoms
\begin{equation}
n=\frac{1}{L^3}
\end{equation}

\subsubsection{The velocity of zero-point oscillations of helium atom}
It is naturally  to suppose that zero-point oscillations of atoms are  harmonic  and the equality of kinetic and potential energies are characteristic for them:
\begin{equation}
\frac{M_4 \widehat{v_0}^2}{2} - \frac{e^2}{a_B}a_B^6 n^2 = 0,
\end{equation}
where $M_4$ is mass of helium atom, $\widehat{v_0}$ is their averaged velocity of harmonic zero-point oscillations.\\

Hence, after simple transformations we obtain:
\begin{equation}
\widehat{v_0}=c\alpha^3\left\{\frac{n}{n_0}\right\},\label{Ea31}
\end{equation}
where the notation is introduced:
\begin{equation}
n_0=\frac{\alpha^2}{a_B^3}\sqrt{\frac{M_4}{2m_e}}.\label{}
\end{equation}
If the expression in the curly brackets
\begin{equation}
\frac{n}{n_0}=1,\label{edin}
\end{equation}
we obtain
\begin{equation}
\widehat{v_0}=c\alpha^3\cong 116.5~
m/s.\label{alpha3}
\end{equation}

\subsubsection{The density of liquid helium}
The condition (\ref{edin}) can be considered as the definition of the density of helium atoms in the superfluid state:
\begin{equation}
n=n_0=\frac{\alpha^2}{a_B^3}\sqrt{\frac{M_4}{2m_e}}\cong 2.172\cdot 10^{22}~ atom/cm^3.\label{nnn}
\end{equation}
According to this definition, the density of liquid helium-4
\begin{equation}
\gamma_4 = n M_4\cong 0.144~ g/cm^3%\label{}
\end{equation}
that is in good agreement with the measured density of the liquid helium $0.145~g/cm^3$ for $T\simeq T_\lambda$.

Similar calculations for liquid helium-3 gives the density  $0.094~g/cm^3$, which can be regarded as consistent with its density $0.082 ~g/cm^3$ experimentally measured near the boiling point.

\subsubsection{The dielectric constant of liquid helium}
To estimate the dielectric constant of helium we can use
the Clausius-Mossotti equation \cite{Fr}:
\begin{equation}
\frac{\varepsilon-1}{\varepsilon+2}=\frac{4\pi}{3}{n}{A}.\label{KM}
\end{equation}
At taking into account Eq.(\ref{vA}), we obtain
\begin{equation}
\varepsilon\approx 1.040,
\end{equation}
that differs slightly from the dielectric constant of the liquid helium,
measured near the $\lambda$-point \cite{Russ}:
\begin{equation}
\varepsilon\approx 1.057
\end{equation}

\subsubsection{The temperature of $\lambda$-point}
The superfluidity is destroyed at the temperature $T_\lambda$, at which the energy of thermal motion is compared with the energy of the Van-der-Waals bond
in superfluid condensate
\begin{equation}
\frac{3}{2}kT_\lambda - \frac{e^2}{a_B}a_B^6 n^2 = 0.
\end{equation}
With taking into account Eq.(\ref{nnn})
\begin{equation}
T_{\lambda}=\frac{1}{3k}\frac{M_4}{m_e}\frac{\alpha^4 e^2}{a_B}\label{f021}
\end{equation}
or after appropriate substitutions
\begin{equation}
T_{\lambda}=\frac{1}{3}\frac{M_4 c^2 \alpha^6}{k} =2.177 K,\label{f022}
\end{equation}
that is in very good agreement with the measured value $T_\lambda = 2.172K$.\footnote{There is a unexpected fact. The expression (\ref{f022}) for the temperature of $\lambda$-transition is given without any explanations in some articles of Internet at citing of patents \cite{Il}.  These articles and patents say nothing at all about zero-point oscillations, and  don't give generally any explanations of the reasons that allowed to write this expression.}

\subsubsection{The boiling temperature of liquid helium}
After comparison of Eq.(\ref{f1}) - Eq.(\ref{f2}), we have
\begin{equation}
T_{boil}=2T_\lambda=4.35 K\label{Tboil}
\end{equation}
This is the basis for the assumption that the liquefaction of helium is due to the attractive forces between the atoms with ordered
lengthwise  components of their oscillations.

\subsubsection{The velocity of the first sound in liquid helium}
It  is known from the theory of the harmonic oscillator that the maximum value of its velocity  is twice bigger than its average velocity.
In this connection, at assumption that the first sound speed  $c_{s1}$ is limited by this maximum speed oscillator, we obtain
\begin{equation}
c_{s1}=2\widehat{v_0}\simeq 233~m/s.\label{v01}
\end{equation}
It is in consistent with the measured value of the velocity of the first sound in helium, which has the maximum value of $238.3~m/s$ at $T\rightarrow 0$  and decreases with increasing temperature up to about $220~m/s$ at $T=T_\lambda$.\\

%\newpage
The results obtained in this subsection  are summarized  for clarity in the Table.(\ref{DT}).\\

The measurement data in this table are mainly quoted by \cite{Kik} and \cite{Russ}.\\

%\newpage
{
{
\begin{table}
\centering
\begin{tabular}{||c|c|c||c||}\hline\hline
&&&\\%\hline
  &defining &calculated&measured\\
  parameter&&&\\
  &formula&value&value\\
  &&&\\\hline
  the velocity of zero-point&&&\\
  oscillations of&$\widehat{v_0}=c\alpha^3$&$116.5$~&\\
  helium atom &&m/s&\\\hline\hline
  The density of atoms&&&\\
  in liquid &$n=\sqrt{\frac{M_4}{2m_e}}\frac{\alpha^2}{a_B^3}$&$2.172\cdot 10^{22}$&\\
  helium &&$atom/cm^3$&\\\hline\hline
  The density&&&\\
  of liquid helium-4&$\gamma=M_4 n$&$144.3$&$145_{T\simeq T_{\lambda}}$\\
   $ g/l$&&&\\\hline\hline
  The dielectric&&&$1.048_{T\simeq 4.2}$\\
  constant&$\frac{\varepsilon-1}{\varepsilon+2}=\frac{4\pi}{3}{\alpha^2}{\sqrt{\frac{M_4}{2m_e}}}$&1.040&\\
  of liquid helium-4&&&$1.057_{T\simeq T_{\lambda}}$\\\hline\hline
  The temperature &&&\\
  &$T_\lambda\simeq\frac{M_4 c^2 \alpha^6}{3}$&$2.177$&$2.172$\\
  $\lambda$-point,K&&&\\\hline\hline
  The boiling &&&\\
  temperature&$T_{boil}\simeq 2T_\lambda$&$4.35$&$4.21$\\
  of helium-4,K&&&\\\hline\hline
  The first sound &&&\\
  velocity,&$c_{s1}=2\widehat{v_0}$&$233$&$238.3_{T\rightarrow 0}$\\
  $m/s$&&&\\\hline\hline
\end{tabular}
\caption{Comparison of the calculated values of liquid helium-4 with the measurement data}
\label{DT}
\end{table}
}}
\newpage
\subsubsection{The estimation of characteristic properties of He-3}
The results of similar calculations for the helium-3 properties are summarized in the Tab.(\ref{DT3}).
\bigskip

{
{\begin{table}
\centering
\hspace{-1cm}\begin{tabular}{||c|c|c||c||}\hline\hline
&&&\\%\hline
 &defining &calculated&measured\\
  parameter&&&\\
  &formula&value&value\\
  &&&\\\hline
  The velocity of zero-point&&&\\
  oscillations of&$\widehat{v_0}=c\alpha^3$&$116.5$&\\
  helium atom &&m/s&\\\hline\hline
  The density of atoms&&&\\
  in liquid &$n_3=\sqrt{\frac{M_3}{2m_e}}\frac{\alpha^2}{a_B^3}$&$1.88\cdot 10^{22}$&\\
  helium-3 &&$atom/cm^3$&\\\hline\hline
  The density&&&\\
  of liquid&$\gamma=M_3 n_3$&$93.7~$&$82.3~$\\
  helium-3, g/l&&&\\\hline\hline
  The dielectric&&&\\
  constant&$\frac{\varepsilon-1}{\varepsilon+2}=\frac{4\pi}{3}{\alpha^2}{\sqrt{\frac{M_3}{2m_e}}}$&1.035&\\
  of liquid helium-3&&&\\\hline\hline
   The boiling&&&\\
   temperature&$T_{boil}\simeq \frac{4}{3}\frac{\mathcal{E}_{W}}{k}$&$3.27$&$3.19$\\
  of helium-3,K&&&\\\hline\hline
   The sound velocity&&&\\
  in liquid&$c_{s}=2\widehat{v_0}$&233~&\\
  helium-3&&m/s&\\\hline\hline
  \end{tabular}
  \caption{The characteristic properties of liquid helium-3}
\label{DT3}
\end{table}

\bigskip

There is a radical difference between mechanisms of transition to the superfluid state for He-3 and He-4.
Superfluidity occurs if complete ordering exists in the atomic system.
For superfluidity of He-3 electromagnetic interaction should order not only zero-point vibrations of atoms, but also the magnetic moments of the nuclei.

It is important to note that  all characteristic dimensions of this task: the amplitude of the zero-point oscillations, the atomic radius, the distance between  atoms in liquid helium - all equal to the Bohr radius $a_B$ by the order of magnitude.  Due to this fact, we can estimate the oscillating magnetic field, which a fluctuating electronic shell creates  on "its" \ nucleus:
\begin{equation}
H_\Omega\approx\frac{e}{a_B^2}\frac{a_B \Omega_0}{c}\approx \frac{\mu_B}{A_3},
\end{equation}
where $\mu_B=\frac{e\hbar}{2m_ec}$ is the Bohr magneton, $A_3$ is the electric polarizability of helium-3 atom.

Because the value of magnetic moments for the nuclei He-3 is approximately equal to the nuclear Bohr magneton $\mu_{n_B}=\frac{e\hbar}{2m_pc}$, the ordering in their system must occur below the critical temperature
\begin{equation}
T_c = \frac{\mu_{n_B}H_\Omega}{k}\approx  10^{-3} K.%.\label{f0}
\end{equation}
This finding is in agreement with the measurement data.
The fact that the nuclear moments can be arranged in parallel or antiparallel to each other is consistent with the presence of the respective phases of superfluid helium-3.\\

Concluding this approach permits to explain the mechanism of superfluidity in liquid helium.

In this way, the apposite  quantitative estimations of main parameters of the liquid helium and its transition to the superfluid state were  obtained.

 It was established that both related phenomena, superconductivity and  superfluidity, are based on the same physical mechanism: they both are consequences of the ordering of  zero-point oscillations.

%\newpage

\section{Conclusion}
Until now it has been commonly thought that the existence of the isotope effect in superconductors leaves only one way for explanation  of the superconductivity phenomenon - the way based on the phonon mechanism.

Over fifty years of theory development based on the phonon mechanism, has not lead to success. All attempts to explain why some superconductors have certain critical temperatures (and critical magnetic fields) have failed.

This problem was further exacerbated with the discovery of high temperature superconductors. How can we move forward in HTSC understanding, if we cannot understand the mechanism that determines the critical temperature elementary superconductors?

 In recent decades, experimenters have shown that  isotopic substitution in metals leads to a change in the parameters of their crystal lattice and thereby affect the Fermi energy of the metal. As results, the  superconductivity can be based on a nonphonon mechanism.

The theory proposed in this paper suggests that the specificity of the association mechanism of electrons pairing is not essential.
It is merely important that such a mechanism was operational over the whole considered range of temperatures. The nature of the mechanism forming the electron pairs  does not matter, because although the work of this mechanism  is necessary it is still not a sufficient condition for the superconducting condensate's existence.
This is caused by the fact that after the electron pairing, they still remain as non-identical particles and  cannot form the condensate, because the individual pairs differ from each other as they commit uncorrelated zero-point oscillations.
 Only after an ordering of these zero-point oscillations, an energetically favorable lowering of the energy can be reached and a condensate at the level of minimum energy can then be formed.
Due to this reason the  ordering of  zero-point oscillations must be considered as the cause of the occurrence of superconductivity.

Therefore, the density of superconducting carriers and the critical temperature of a superconductor are determined by the Fermi energy of the metal, The critical magnetic field of a superconductor is given by the mechanism of destruction of the coherence of zero-point oscillations.

In conclusion, the consideration of zero-point oscillations allows us  to construct the theory of superconductivity, which is characterized by the ability to give estimations for the critical parameters of elementary superconductors. These results  are in satisfactory agreement with measured data.

This approach permit to explain the mechanism of superfluidity in liquid helium.
For electron shells  of atoms in S-states, the energy of interaction of zero-point oscillations  can be considered as a manifestation of  Van-der-Waals forces.
In this way the apposite  quantitative estimations of temperatures of the helium liquefaction  and its transition to the superfluid state was  obtained.

Thus it is established that both related phenomena, superconductivity and  superfluidity, are based on the same physical mechanism - they both are consequences of the ordering of  zero-point oscillations.

%\newpage

\end{document}